\documentclass{aa}  
\usepackage{graphicx}
\usepackage[varg]{txfonts}
\usepackage{hyperref}
\usepackage{xcolor, subcaption, textcomp, lscape, afterpage}
\newcommand\B{\rule[-1.2ex]{0pt}{0pt}} 
\begin{document} 

   \title{Low NH$_{3}$/H$_{2}$O ratio in comet C/2020 F3 (NEOWISE) at $0.7$~au from the Sun}
	 \titlerunning{NH$_{3}$/H$_{2}$O in comet C/2020 F3 (NEOWISE)}

   \author{Maria N. Drozdovskaya\inst{1}
          \and
					Dominique Bockel\'{e}e-Morvan\inst{2}
					\and
					Jacques Crovisier\inst{2}
					\and
					Brett A. McGuire\inst{3, 4}
					\and\\
          Nicolas Biver\inst{2}
					\and
					Steven B. Charnley\inst{5}
					\and
					Martin A. Cordiner\inst{5, 6}
					\and
					Stefanie N. Milam\inst{5}
					\and
					Cyrielle Opitom\inst{7}
					\and
					Anthony J. Remijan\inst{4}
          }

   \institute{Center for Space and Habitability, Universit\"{a}t Bern, Gesellschaftsstrasse 6, CH-3012 Bern, Switzerland\\
              \email{maria.drozdovskaya@unibe.ch; maria.drozdovskaya.space@gmail.com}
         \and
				     LESIA, Observatoire de Paris, Universit\'{e} PSL, Sorbonne Universit\'{e}, Universit\'{e}  Paris Cit\'{e}, CNRS, 5 place Jules Janssen, 92195 Meudon, France\\
						 \email{Dominique.Bockelee@obspm.fr}
				 \and
				     Department of Chemistry, Massachusetts Institute of Technology, Cambridge, MA 02139, USA
				 \and
						 National Radio Astronomy Observatory, Charlottesville, VA 22903, USA
				 \and
				     Astrochemistry Laboratory, NASA Goddard Space Flight Center, 8800 Greenbelt Road, Greenbelt, MD 20771, USA
				 \and
						 Department of Physics, Catholic University of America, Washington, DC 20064, USA
				 \and
             Institute for Astronomy, University of Edinburgh, Royal Observatory, Edinburgh EH9 3HJ, UK
             }

   \date{Received 14 March 2023; accepted 19 July 2023}

 
  \abstract
   {A lower-than-solar elemental nitrogen content has been demonstrated for several comets, including 1P/Halley and 67P/Churyumov-Gerasimenko (67P/C-G) with independent in situ measurements of volatile and refractory budgets. The recently discovered semi-refractory ammonium salts in 67P/C-G are thought to be the missing nitrogen reservoir in comets.}
   {The thermal desorption of ammonium salts from cometary dust particles leads to their decomposition into ammonia (NH$_{3}$) and a corresponding acid. The NH$_{3}$/H$_{2}$O ratio is expected to increase with decreasing heliocentric distance with evidence for this in near-infrared observations. NH$_{3}$ has been claimed to be more extended than expected for a nuclear source. Here, the aim is to constrain the NH$_{3}$/H$_{2}$O ratio in comet C/2020 F3 (NEOWISE) during its July 2020 passage.}
   {OH emission from comet C/2020 F3 (NEOWISE) was monitored for 2 months with the Nan\c{c}ay Radio Telescope (NRT) and observed from the Green Bank Telescope (GBT) on 24 July and 11 August 2020. Contemporaneously with the 24 July 2020 OH observations, the NH$_{3}$ hyperfine lines were targeted with GBT. From the data, the OH and NH$_{3}$ production rates were derived directly, and the H$_{2}$O production rate was derived indirectly from the OH.}
   {The concurrent GBT and NRT observations allowed the OH quenching radius to be determined at $\left(5.96\pm0.10\right)\times10^{4}$~km on 24 July 2020, which is important for accurately deriving $Q(\text{OH})$. C/2020 F3 (NEOWISE) was a highly active comet with $Q(\text{H}_{2}\text{O}) \approx 2\times10^{30}$~molec~s$^{-1}$ one day before perihelion. The $3\sigma$ upper limit for $Q_{\text{NH}_{3}}/Q_{\text{H}_{2}\text{O}}$ is $<0.29\%$ at $0.7$~au from the Sun.}
   {The obtained NH$_{3}$/H$_{2}$O ratio is a factor of a few lower than measurements for other comets at such heliocentric distances. The abundance of NH$_{3}$ may vary strongly with time depending on the amount of water-poor dust in the coma. Lifted dust can be heated, fragmented, and super-heated; whereby, ammonium salts, if present, can rapidly thermally disintegrate and modify the $\text{NH}_{3}/\text{H}_{2}\text{O}$ ratio.}
	
   \keywords{astrochemistry -- comet: individual: C/2020 F3 (NEOWISE) -- Radio lines: general -- ISM: molecules}

   \maketitle
%

\section{Introduction}
\label{Introduction}

Comets are ice-rich, kilometer-sized bodies composed of volatiles, refractories, and semi-refractory compounds. The key characteristic of these minor bodies of our Solar System is their outgassing activity, especially upon their approach toward the Sun. During the approach, the volatile ices in a cometary nucleus are released as gases into a tenuous coma and dust particles are lifted off of the surface through several mechanisms \citep{JewittHsieh2022}. Based on their physical properties, it remains unclear if such bodies formed in a protoplanetary disk through hierarchical agglomeration of cometesimals or through pebble accretion \citep{Weissman2020, Blum2022}. The chemical composition of volatiles does support partial inheritance of ices from the earliest stages of star formation to cometary nuclei \citep{Bockelee-Morvan2000, Drozdovskaya2019, Altwegg2019}.

The dominant volatile constituent of comets is water (H$_{2}$O; \citealt{MummaCharnley2011}). A typical, although not only, method for remote sensing of cometary water is through observations of OH \citep{SchloerbGerard1985}, which is a primary product of H$_{2}$O photodissociation in the coma. OH lines are available in the UV \citep{Feldman2004, Bodewits2022}, radio \citep{Crovisier2002a, Crovisier2002b}, and infrared (IR; \citealt{Bonev2006, BonevMumma2006}) regimes. The production rate of H$_{2}$O is a fundamental parameter of a comet that must be constrained for each comet individually and over the longest-possible period of time in order to get a handle on the individual time-dependent cometary activity. In turn, this facilitates comparisons of chemical compositions from comet to comet relative to the dominant volatile species, H$_{2}$O.

The nitrogen budget of comets has been a long-standing mystery, because even summing all the volatile N measured in the coma with the refractory N measured in the dust leaves a deficiency in N relative to the solar N/C ratio \citep{Geiss1988, Rubin2015a, Biver2022a}. The recent discovery of ammonium salts (NH$_{4}^{+}$X$^{-}$) on the dust grains of comet 67P/Churyumov-Gerasimenko (hereafter, 67P/C-G) is thought to have uncovered the missing reservoir of nitrogen in comets \citep{Poch2020, Altwegg2020a}. Their presence was already hypothesized on the basis of data for comet 1P/Halley \citep{Wyckoff1991} and the salt NH$^{+}_{4}$CN$^{-}$ was suggested to be the precursor for NH$_{3}$ and HCN release in more recent comets \citep{Mumma2017, Mumma2018, Mumma2019}. Ammonium salts are considered to be semi-volatile with sublimation temperatures in the $160-230$~K range \citep{Bossa2008b, Danger2011b}. In the laboratory, it was demonstrated that ammonia (NH$_{3}$) is a key product of thermal desorption of ammonium salts that decompose during this process \citep{Haenni2019, Kruczkiewicz2021}. Consequently, it is expected that as the comet's heliocentric distance decreases, its ammonium salts desorb and decompose, which in turns leads to an increasing NH$_{3}$/H$_{2}$O ratio. This trend was demonstrated on the basis of high-resolution IR spectroscopic observations of $30$ comets (fig.~$7$b of \citealt{DelloRusso2016b}). To an extent, this is also seen in the NH$_{3}$/H$_{2}$O ratio measured with the Rosetta Orbiter Spectrometer for Ion and Neutral Analysis (ROSINA; \citealt{Balsiger2007}) instrument aboard the ESA \textit{Rosetta} spacecraft during its 2-yr long monitoring of 67P/C-G, although there is quite a bit of scatter in the ratio that is uncorrelated with the latitude and season (fig.~$1$a of \citealt{Altwegg2020a}).

Ammonium salts have also been detected on the surface of the dwarf planet Ceres with the Dawn mission \citep{DeSanctis2015b, DeSanctis2019}, but they have not been detected in interstellar ices thus far \citep{Boogert2015}. However, their presence has long been hypothesized \citep{LewisPrinn1980} and expected based on a suite of laboratory experiments involving the UV irradiation or electron bombardment of NH$_{3}$-containing ices \citep{Grim1989a, Bernstein1995, MunozCaroSchutte2003, Gerakines2004, vanBroekhuizen2004, Bertin2009, Vinogradoff2011}. It has also been demonstrated in the laboratory that acid-base grain-surface reactions between NH$_{3}$ and HNCO, NH$_{3}$ and HCN, and NH$_{3}$ and HCOOH may form the ammonium cyanate (NH$^{+}_{4}$OCN$^{-}$), ammonium formate (NH$^{+}_{4}$HCOO$^{-}$), and ammonium cyanide (NH$^{+}_{4}$CN$^{-}$) salts, respectively, even at prestellar core temperatures of $\sim10$~K without energetic processing and increase in efficiency with elevated temperatures \citep{Schutte1999, Raunier2003a, Raunier2004b, Galvez2010, Mispelaer2012, Noble2013a, Bergner2016}. At a warmer temperature of $80$~K, NH$_{3}$ and CO$_{2}$ have also been experimentally shown to react to produce ammonium carbamate (NH$^{+}_{4}$H$_{2}$NCOO$^{-}$; \citealt{Bossa2008b}).

Detections of the NH$_{3}$ hyperfine inversion lines, which arise from the oscillation of the nitrogen atom through the plane formed by the three hydrogens, have been challenging to secure on a consistent basis in comets. A possible detection of the $J_K=3_3-3_3$ inversion line at $23.870$~GHz was made in comet C/1983 H1 (IRAS-Araki-Alcock) with the Effelsberg 100-m telescope \citep{Altenhoff1983}. A sensitive search for the same line with the same instrument in comets  1P/Halley and 21P/Giacobini–Zinner was negative \citep{Bird1987}, which was also the case for the earlier comet C/1973 E1 (Kohoutek) \citep{Churchwell1976}. A confirmed detection of the $J_K=1_1-1_1$ and $3_3-3_3$ inversion lines with high signal-to-noise ratios was obtained in C/1996 B2 (Hyakutake) with the 43-m telescope of the National Radio Astronomy Observatory (NRAO) in Green Bank \citep{Palmer1996}. In comet C/1995 O1 (Hale-Bopp), all inversion transitions from $J_K = 1_1-1_1$ to $5_5-5_5$ were detected with the Effelsberg 100-m telescope \citep{Bird1997}. Some of these were also observed with the NRAO 43-m telescope \citep{Butler2002} and the 45-m telescope of the Nobeyama Radio Observatory \citep{Hirota1999}. Despite a sensitive search at the Effelsberg 100-m telescope, NH$_3$ was not detected in comets C/2001 A2 (LINEAR) , C/2001 Q4 (NEAT), and C/2002 T7 (LINEAR), and only marginally detected in comet 153P/Ikeya-Zhang \citep{Bird2002, Hatchell2005a}. Subsequent rotational lines, which are stronger, lie at submillimeter wavelengths. NH$_3$ was detected in several comets (Table~\ref{tabnh3comets}) using the \textit{Odin} and \textit{Herschel} satellites through its $J_K = 1_0-0_0$ transition at $572$~GHz that is only accessible from space \citep{Biver2007b, Biver2012}. This transition was also monitored in comet 67P/C-G with the Microwave Instrument for the Rosetta Orbiter (MIRO) instrument aboard \textit{Rosetta} \citep{Gulkis2007, Biver2019}. At shorter IR wavelengths, NH$_3$ has been observed in many comets (Table~\ref{tabnh3comets}) through its vibrational lines near $3~\mu$m, although its unblended lines are sparse and weak in comparison to other species typically observed in the near-IR (\citealt{DelloRusso2016b, DelloRusso2022, Bonev2021} and the many references therein).

In this paper, we present radio observations of comet C/2020~F3~(NEOWISE), hereafter F3 for simplicity, taken at the Nan\c{c}ay Radio Telescope (NRT) and the Green Bank Telescope (GBT) that targeted OH and NH$_{3}$ lines (Table~\ref{table:lines}). The long-period comet F3 originating from the Oort Cloud was a bright naked-eye object in the sky that underwent perihelion on 3 July 2020 and perigee on 22-23 July 2020. A more detailed description of the comet is given in Section~\ref{sec:comet}. The OH emission from the comet was monitored nearly continuously for 2 months at the NRT pre- and post-perihelion. GBT observations targeted contemporaneously OH and NH$_{3}$ lines post-perihelion on 24 July 2020. Additional OH data were collected at the GBT on 11 August 2020. The GBT observational strategy and data reduction are presented in Section~\ref{GBT}. The NRT observations are described in Section~\ref{NRT}. The methods behind the determination of the OH production rate are described in Section~\ref{OHprodrate}. The analysis of OH line profiles in the context of outflow velocities is in Section~\ref{sec:lineprofile}. GBT and NRT OH observations taken on the same date, a mere $2$~hours apart, are used to constrain the OH quenching radius in Section~\ref{sec:combined-OH-analysis} based directly on observational data, which is critical for an accurate determination of the OH production rate. The GBT OH observations of 11 August 2020 are analyzed in the context of the Greenstein effect in Section~\ref{GBTAug11}. The OH and H$_{2}$O production rates are finally computed in Section~\ref{sec:NRT-OHprod}. GBT NH$_{3}$ observations are analyzed in Section~\ref{sec:NH3} and discussed in the context of the NH$_{3}$/H$_{2}$O ratio as a function of heliocentric distance in Section~\ref{discussion}. The conclusions of the work can be found in Section~\ref{conclusions}.

\begin{table}
\caption{Targeted lines.}
\label{table:lines} 
\begin{center}
\begin{tabular}{lr}
\hline\hline\noalign{\vskip 2mm}
Line & Frequency \\
& (MHz) \\
\hline\noalign{\vskip 2mm}
OH $^2\Pi_{3/2}~J=3/2~F=1-2$  & $1~612.2309$ \\
OH $^2\Pi_{3/2}~J=3/2~F=1-1$  & $1~665.4018$ \\
OH $^2\Pi_{3/2}~J=3/2~F=2-2$  & $1~667.3590$ \\
OH $^2\Pi_{3/2}~J=3/2~F=2-1$  & $1~720.5299$ \\
NH$_3$ $J_K=1_1-1_1$ & $23~694.4955$ \\
NH$_3$ $J_K=2_2-2_2$ & $23~722.6333$  \\
NH$_3$ $J_K=3_3-3_3$ & $23~870.1292$ \\
NH$_3$ $J_K=4_4-4_4$ & $24~139.4163$  \\
NH$_3$ $J_K=5_5-5_5$ & $24~532.9887$ \\
\hline
\end{tabular}
\end{center}
\end{table}


\section{Observations of comet C/2020 F3 (NEOWISE)}
\label{observations}

\subsection{Comet C/2020 F3 (NEOWISE)}
\label{sec:comet}
\begin{figure}
\includegraphics[width=\hsize]{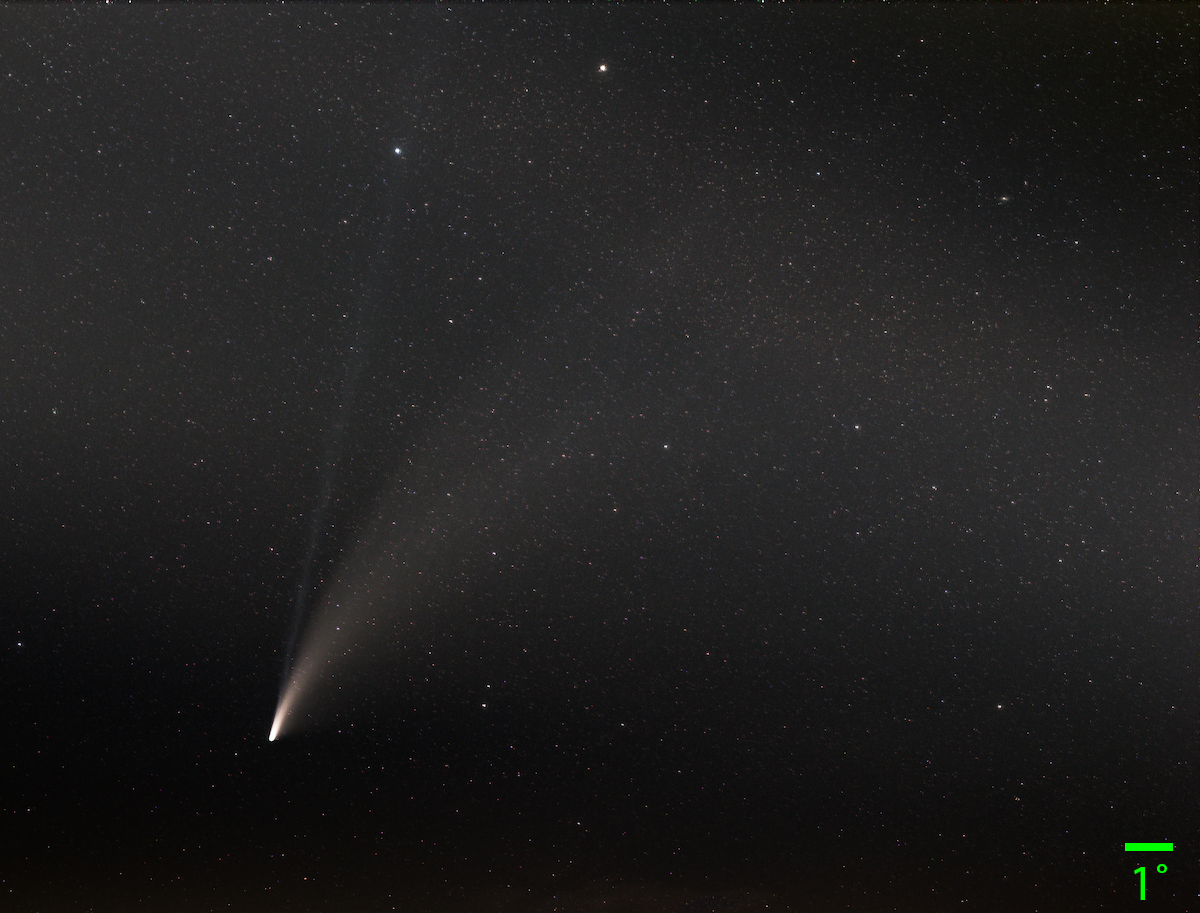}
\caption{Comet C/2020 F3 (NEOWISE) as seen from the Blue Ridge Mountains outside of White Hall, Virginia on 20 July 2020. The image is an aligned and stacked average of five, 5-minute exposures taken using a Nikon D750 DLSR camera with a $50$~mm $f/1.8$ lens held on an equatorial tracking mount (B. A. McGuire). The image is approximately $29.3^{\circ} \times 22.3^{\circ}$ in total. The $1^{\circ}$ scale is indicated in the lower right corner, which corresponds to $3\times10^{4}$~km at the geocentric distance of the comet on this date. A finder chart for this image is available at \url{https://nova.astrometry.net/user_images/7997658\#} \citep{Lang2010}.}
\label{fig:neowise}
\end{figure}

Comet F3 is considered to be the brightest comet in the northern hemisphere since comet C/1995~O1~(Hale-Bopp) in 1997 (Fig.~\ref{fig:neowise}). It was a bright naked-eye target in midsummer of 2020 and termed by some as the ``Great Comet of 2020''. Comet F3 showed a huge ($\sim10^{5}$~km), curving, banded dust tail and a much fainter, straight, blue ion tail ($\sim10^{6}$~km). It was viewed from the International Space Station (ISS)\footnote{Observed on 5 July 2020 and viewed at \url{https://science.nasa.gov/comet-neowise-iss} on 27 June 2023.]} and photographed by countless members of the public\footnote{For example, even winning first place in the 2021 IAU OAE Astrophotography Contest in the Comets category: ``Neowise's metamorphosis'' by Tom\`{a}\v{s} Slovinsk\'{y} and Petr Hor\'{a}lek (Slovakia) spanning observations from 9 July to 2 August 2020 and viewed at \url{https://www.iau.org/public/images/detail/ann21047d/} on 27 June 2023.}. F3 is a long-period comet originating from the Oort Cloud with an orbital period of $\sim4~500$~yr and a near-parabolic (eccentricity of $0.999$) trajectory. It was discovered on March 27th, 2020 by the Near Earth Object Wide-field Infrared Survey Explorer (NEOWISE) mission of the Wide-field Infrared Survey Explorer spacecraft (Mainzer et al. 2011)\footnote{Minor Planet Electronic Circular (MPEC) 2020-G05: \url{https://minorplanetcenter.net/mpec/K20/K20G05.html}.}. Perihelion took place on July 3rd, 2020 at $0.295$~au from the Sun and perigee occurred in the night from July 22nd to 23rd, 2020 at $0.692$~au from the Earth. Its nucleus was estimated to be approximately $5$~km in diameter \citep{Bauer2020} with a rotation period of $7.8 \pm 0.2$~h \citep{Manzini2021}. Sudden short-lived outbursts have not been reported for comet F3 and none were observed during the extended (from May to September 2020) monitoring campaign of the comet by the Solar Wind ANisotropies (SWAN) camera on the Solar and Heliospheric Observatory (SOHO) spacecraft \citep{Combi2021}. However, strong jet activity is seen in the HST images taken on 8 August 2020 (Program ID: 16418, P.I.: Qicheng Zhang)\footnote{Observed on 8 August 2020 and seen at \url{https://hubblesite.org/contents/media/images/2020/45/4731-Image?news=true} on 27 June 2023.}. There were also no signs of disintegration pre-perihelion based on NASA/ESA's Solar and Heliospheric Observatory (SOHO) LASCO C3 coronagraph observations \citep{KnightBattams2020}.

Comet F3 was observed by several facilities in the optical regime. Observations with the High Accuracy Radial velocity Planet Searcher for the Northern hemisphere (HARPS-N) echelle spectrograph on the 360-cm Telescopio Nazionale Galileo (TNG) with a high resolving power ($R$) of $115~000$ revealed detections of C$_{2}$, C$_{3}$, CN, CH, NH$_{2}$, Na, and [O~I] (taken on 26 July and 5 August 2020; \citealt{Cambianica2021}). Lines of C$_{2}$, NH$_{2}$, Na, and [O~I] were also detected with low resolution, $R=7~400-9~300$, observations made with the \'{E}chelle spectrograph FLECHAS at the 90-cm telescope of the University Observatory Jena taken on 21, 23, 29, 30, and 31 July 2020 (\citealt{BischoffMugrauer2021}; Na was detected only on the first two dates). Lines of C$_{2}$, C$_{3}$, CN, CH, NH$_{2}$, Na, and [O~I] were also detected with the Tull Coud\'{e} Spectrograph at $R=60~000$ on the 2.7-m Harlan J. Smith Telescope at McDonald Observatory on 24 July and 11 August 2020 \citep{Cochran2020}. Lines of C$_{2}$, C$_{3}$, CN, NH$_{2}$, Na, [O~I], K, and H$_{2}$O$^{+}$ were also detected with ultra-high resolution spectroscopy at $R=140~000$ with the EXtreme PREcision Spectrograph (EXPRES) on the 4.3-m Lowell Discovery Telescope on 15 and 16 July 2020 \citep{Ye2020}. Further chemical characterization of comet F3 was executed in the IR. In the $1.1-5.3$~$\mu$m range, data were taken with the long-slit near-IR high-resolution ($R=35~000-70~000$) immersion echelle spectrograph iSHELL at the NASA/IRTF facility. The $0.95-5.5$~$\mu$m range was observed with NIRSPEC~2.0 ($R=25~000-37~500$) at the Keck Observatory. These near-IR observations of \citet{Faggi2021} led to the detection of $9$ primary volatiles (H$_{2}$O, HCN, NH$_{3}$, CO, C$_{2}$H$_{2}$, C$_{2}$H$_{6}$, CH$_{4}$, CH$_{3}$OH, and H$_{2}$CO) and $3$ product species (CN, NH$_{2}$, OH$^{*}$) on several dates in July and August 2020. In the radio, comet F3 was observed with the Institut de Radio Astronomie Millim\'{e}trique (IRAM)~30-m and the NOrthern Extended Millimeter Array (NOEMA) telescopes on several dates in July and August with secured detections of HCN, HNC, CH$_{3}$OH, CS, H$_{2}$CO, CH$_{3}$CN, H$_{2}$S, and CO \citep{Biver2022b}.

\subsection{Green Bank Telescope observations}
\label{GBT}

Comet F3 was observed with the Robert C. Byrd 100-m GBT in Green Bank, West Virginia under the project code GBT20A-587. The Director's Discretionary Time (DDT) proposal was awarded a total amount of $12$~h, which was to be executed in two $6$~h sessions with a separation of at least $3$ days to probe dependencies with heliocentric distance. It was planned for each session to begin with $1$~h of OH observations, followed immediately by $5$~h of NH$_{3}$ observations (including $5$ and $40$~min of overheads, respectively). The targeted lines are tabulated in Table~\ref{table:lines}. Due to an unfortunate overlap of the comet's perihelion on 3 July 2020 with scheduled maintenance, both sessions were observed post-perihelion.

\begin{table}
\caption{GBT and NRT half-power beam widths (HPBWs) and efficiencies.}
\label{table:GBT}
\begin{center}
\begin{tabular}{clccc}
\hline\hline\noalign{\vskip 2mm}
Facility & Frequency & $\theta_{\text{HPBW}}$ & $\eta_{\text{A}}$ & $\eta_{\text{mB}}$\\
  & (GHz) & ($\arcmin$) & &\\ 
\hline\noalign{\vskip 2mm}
GBT & $1.67$ & $7.55\pm0.083$ & $0.72^a$ & $0.95^b$\\
GBT & $23.8$ & $0.53\pm0.005$ & $0.68^a$ & $0.91^b$\\
\hline
NRT & $1.67$ & $3.5\times19$ & $0.46$ & $0.65^c$\\
\hline
\end{tabular}
\end{center}
\footnotesize{$^{a}$ Aperture efficiency ($\eta_{\text{A}}$) from the GBO Proposer's Guide for the GBT of 3 January 2023.\\
$^b$ Main beam efficiency deduced from $\eta_{\text{mB}} = \eta_{\text{A}} S_{\text{tel}} \Omega/\lambda^{2}$, where $S_{\text{tel}}$ is the GBT surface telescope area ($7~854$~m$^{2}$), $\lambda$ is the wavelength, and $\Omega$ is the solid angle of the Gaussian beam defined by the HPBW in radians, which is given by $\Omega = \pi \theta_{\text{HPBW}}^{2} / 4 \ln 2$.} \\
$^c$ Assuming a conversion factor of $0.85$~K~Jy$^{-1}$ and an effective area of $7~000$~m$^{2}$.
\end{table}

\subsubsection{Day 1 and 2 OH observations}
\label{GBT12OH}

OH observations were carried out with the L-band receiver coupled with the VErsatile GBT Astronomical Spectrometer (VEGAS) backend \citep{Roshi2011}. For the first session, on day 1 (24.71 July 2020), mode $8$ was used with a $100$~MHz bandwidth, $65~536$ channels, and a spectral resolution of $1.5$~kHz ($0.3$~km~s$^{-1}$ at $1~666.0$~MHz). These were single beam, dual polarization observations carried out with in-band frequency switching with a throw of $25$~MHz (centered on the rest frequency of $1~666.0$~MHz). For the second session, on day 2 (11.88 August 2020), the signal was routed to four banks of VEGAS in four different modes: mode $8$ (same as day 1), mode $9$ ($100$~MHz bandwidth, $131~072$ channels, $0.8$~kHz or $0.14$~km~s$^{-1}$ spectral resolution), mode $11$ ($23.44$~MHz bandwidth, $65~536$ channels, $0.4$~kHz or $0.07$~km~s$^{-1}$ spectral resolution), and mode $12$ ($23.44$~MHz bandwidth, $131~072$ channels, $0.2$~kHz or $0.04$~km~s$^{-1}$ spectral resolution). At the start of observing on both days, pointing and focus were performed using the Digital Continuum Receiver (DCR) backend toward the calibrator source 1011+4628 across a $80$~MHz bandwidth. The half-power beam width (HPBW) of the GBT at $1.67$~GHz is $(7.55\pm0.083)\arcmin$ (according to equation $2$ of the Green Bank Observatory (GBO) Proposer's Guide for the GBT of 3 January 2023).

Day 1 OH observations were executed starting 24.71 July 2020. The total integration time on-source was $46$~min ($23$ scans, $2$~min each). Day 2 observations were taken starting 11.88 August 2020. Due to the separation between the two sessions being $17$ days (much longer than the initially envisioned separation of not much more than $3$ days) and the non-detection of NH$_{3}$ at the more favorable smaller heliocentric distance of Day 1 (Section~\ref{GBT1NH3}), the remaining allocated time was dedicated solely to OH observations. However, this time, the observations were performed toward three different locations: on-comet, an offset toward the Sun, and an offset away from the Sun. The integration time was $1$~h~$26$~min ($43$ scans, $2$~min each) on-comet, $1$~h~$38$~min ($49$ scans, $2$~min each) toward-offset, and $1$~h~$58$~min ($59$ scans, $2$~min each) away-offset. Total on-source time was $5$~h~$2$~min. It was envisioned to probe one-beam offsets from the on-comet position along the Sun-comet trajectory to accurately quantify the quenching radius of OH (analogously to the methods of \citet{Colom1999} for comet C/1995 O1 (Hale-Bopp)). However, due to an error in offset coordinates in the observing script, the probed positions were $\sim8$ beams away in RA to the northeast (and $<1/5$ of a beam away in Dec to the southwest). Specifically, the toward-offset was RA, Dec (J2000) of -7m30s, +1$\arcmin$20$\arcsec$ and the away-offset was RA, Dec (J2000) of +7m30s, -1$\arcmin$20$\arcsec$.

The OH $1~665$ and $1~667$~MHz spectra obtained on Day 1 (24.71 July 2020) at the GBT are shown in Fig.~\ref{fig:trapeziumfitting}. The OH $1~667$ MHz spectrum obtained on Day 2 (11.88 August 2020) at the GBT is shown in Fig.~\ref{fig:OH-day2}. L-band Day 1 data have been checked for the presence of the $^{18}$OH line at $1~639.5$~MHz (the strongest of three $^{18}$OH lines in range), but it was not detected. The observed frequency range also covered one line of $^{17}$OH, but based on neither the isotopic ratio of $^{18}\text{O}/^{17}\text{O}=3.6$ in the local interstellar medium (ISM; \citealt{Wilson1999}) nor the measured $^{18}\text{O}/^{17}\text{O}=5.3\pm0.4$ in water of comet 67P/C-G \citep{Schroeder2019, MuellerDaniel2022}, it is not expected to be detected if $^{18}$OH is not. The main beam efficiencies ($\eta_{\text{mB}}$) given in Table~\ref{table:GBT} were used to convert the line areas into the main beam brightness temperature scale through $T_{\text{mB}}=T_{\text{A}}^{*}/\eta_{\text{mB}}$. Table~\ref{table:linearea} gives line areas in $T_{\text{A}}^{*}$ scale for the OH lines observed at the GBT. The ratio of the line areas of the $1~667$ and $1~665$~MHz OH lines observed on 24 July 2020 is $1.93 \pm 0.07$, which is consistent, within $2\sigma$, with the statistical ratio of $1.8$.

\subsubsection{Day 1 NH$_{3}$ observations}
\label{GBT1NH3}

NH$_{3}$ observations were carried out on Day 1 (starting 24.81 July 2020). The $7$-pixel K-band Focal Plane Array (KFPA) receiver \citep{Masters2011} was employed with the VEGAS backend. The total integration time on-source was $3$~h~$34$~min ($107$ scans, $2$~min each). Mode $24$ was used with a $23.44$~MHz bandwidth, $65~536$ channels, and a spectral resolution of $0.4$~kHz ($0.005$~km~s$^{-1}$ at $23.8$~GHz). All seven beams of the KFPA were used and routed to six banks of VEGAS  (centered on rest frequencies of $23~694.50$, $23~722.63$, $23~870.13$, $24~139.42$, $24~532.99$, $23~963.9$~MHz). These were dual polarization observations carried out with in-band frequency switching with a throw of $5$~MHz. At the start of observing, pointing and focus were performed using a single spectral window (still with the VEGAS backend) toward the calibrator source 1033+4116 across a $1~500$~MHz bandwidth (centered on the rest frequency of $24~010$~MHz). The HPBW of the GBT at $23.8$~GHz is $(0.53\pm0.005)\arcmin$ (Table~\ref{table:GBT}). The data from all seven beams were inspected for the presence of NH$_{3}$ lines individually; however, solely the centermost on-comet beam has been used for the derivation of the upper limits. Table~\ref{table:linearea} gives the upper limits in $T_{\text{A}}^{*}$ scale on the NH$_{3}$ lines observed at the GBT.

\subsubsection{GBT data reduction}
\label{GBTdatared}

Initial data processing and calibration were performed using \textsc{GBTIDL}\footnote{\textsc{GBTIDL} is an interactive package for reduction and analysis of spectral line data taken with the GBT. See \url{http://gbtidl.nrao.edu/}.}. Day 1 OH and NH$_{3}$ observations were visually inspected on a scan-by-scan basis. All scans were considered for the final data products with folding performed using the \texttt{getfs} routine. The final spectra are noise-weighted averages of all the scans and both polarizations for the case of L-band observations. For the KFPA data, only the right polarization has been considered given a known and reported issue with the left polarization at the time these observations were taken. Day 2 OH observations suffered from much more severe Radio Frequency Interference (RFI). Furthermore, a mishap occurred for modes $11$ and $12$, because the adopted $25$~MHz throw exceeded the bandwidth, resulting in there being no overlap between the sig and ref phases. All $151$ scans were visually inspected in the sig and ref phases (for modes $8$ and $9$; and only in the sig phase for modes $11$ and $12$) separately in both polarizations in small regions near the two OH lines. Scans with severe RFI or odd baselines (even if in just one polarization of either of the phases) were discarded. All scans in the ref phase of modes $8$ and $9$ exhibit an RFI close to $(1667.36-25.00)=1642.36$~MHz, which after folding results in an absorption artifact close to the $1~667.36$~MHz OH line. Consequently, all data in the ref phase from all $4$ modes were discarded in the analysis of the $1~667.36$~MHz OH line. As a result, calibration of the sig phase had to be done manually on the basis of a line- and RFI-free region, and the application of the noise diode in K factor. With this methodology, any remaining absorption feature must be a real signal rather than an artifact of folding. In the analysis of the $1~665.40$~MHz OH line, both sig and ref phases were considered from modes $8$ and $9$ with folding using the \texttt{getfs} routine, but only the sig phase from modes $11$ and $12$ was considered with the manual calibration procedure. Finally, for every mode a noise-weighted average was produced. These spectra were then resampled to the coarsest spectral resolution of mode $8$ ($1.5$~kHz). The resampling has been performed with the Flux Conserving Resampler from the specutils Python package, which conserves the flux during the resampling process \citep{Carnall2017}\footnote{\url{https://specutils.readthedocs.io/en/stable/api/specutils.manipulation.FluxConservingResampler.html\#specutils.manipulation.FluxConservingResampler}}. The final spectrum per OH line per position is a noise-weighted average of the four modes.

For the Day 1 OH data, baseline subtraction was performed (separately for each line) on a scan-by-scan basis with a first-order polynomial in a region centered on the rest-frequency of the specific OH line, which corresponds to $\pm54$~km~s$^{-1}$ ($\pm300$~kHz or $\pm200$~channels), while excluding the central $\pm2$~km~s$^{-1}$ ($\pm11$~kHz or $\pm8$~channels) range that includes the line. For the NH$_{3}$ data, baseline subtraction was performed (separately for each line) on a scan-by-scan basis with a first-order polynomial in a region centered on the rest-frequency of the specific NH$_{3}$ line, which corresponds to $\pm8$~km~s$^{-1}$ ($\pm633$~kHz or $\pm1766$~channels), while excluding the central $\pm2$~km~s$^{-1}$ ($\pm158$~kHz or $\pm442$~channels) range that includes the line. For the Day 2 OH data, baseline subtraction was performed on a scan-by-scan basis with a first-order polynomial in a $0.54-0.76$ (depending on the mode) and $0.2$~MHz wide region near the rest-frequencies of the $1~665.40$ and $1~667.36$~MHz OH lines, respectively. The baseline subtraction was performed solely on the sig phase for the cases when it was the only phase used for the final product (all modes for the $1~667.36$~MHz line, and modes $11$ and $12$ for the $1~665.40$~MHz line). The observed spectra were corrected by the comet's velocity on a scan-by-scan basis based on computations (with a step size of $1$~min) from NASA/JPL Horizons On-Line Ephemeris System (EOP files eop.210205.p210429 and eop.210211.p210505)\footnote{Giorgini, JD and JPL Solar System Dynamics Group, NASA/JPL Horizons On-Line Ephemeris System, \url{https://ssd.jpl.nasa.gov/horizons/}, data retrieved 8 and 12 February 2021 \citep{Giorgini1996, Giorgini2001}, solution JPL\#23.}.

The system temperature ($T_{\text{sys}}$) on Day 1 in the L-band was in the $16.76-19.41$~K range with a mean of $18.01$~K. $T_{\text{sys}}$ on Day 2 in the L-band was in the $11.14-25.79$, $8.77-27.91$, and $17.36-28.14$~K ranges with means of $20.37$, $20.21$, and $21.65$~K at the on-comet, toward-offset, and away-offset positions, respectively. $T_{\text{sys}}$ on Day 1 for the KFPA was in the $69.55-88.29$~K range with a mean of $76.43$~K (in the central beam of the KFPA), which was on the higher end of expectations. The anticipated RMS based on the GBT Sensitivity Calculator was estimated to be in the $2.7-5.4$~mK~km~s$^{-1}$ range for $T_{\text{sys}}\in\left[50-100\right]$~K in the proposal for the KFPA for a line width of $3$~km~s$^{-1}$. However, the attained RMS of the observations came out to be $46-48$~mK~channel$^{-1}$ (in the central beam of the KFPA), which corresponds to $5.6-5.9$~mK~km~s$^{-1}$ for a line width of $3$~km~s$^{-1}$ (using $\text{RMS} (\text{mK~km~s}^{-1}) = \sqrt n \times \text{RMS} (\text{mK~channel}^{-1}) \times dv$, where $dv$ is the channel width and $n$ is the number of channels spanning the line). Other KFPA beams had comparable $T_{\text{sys}}$ and RMS values, except for the sixth beam, which was a factor of $1.7$ higher. The somewhat higher attained RMS in comparison to the anticipated RMS from the GBT Sensitivity Calculator stems from the actual on-source time being $46$~min shorter than planned and the $T_{\text{sys}}$ values being in the higher range of expectations. All spectra are in terms of the antenna temperature corrected for antenna and atmospheric losses, $T_{\text{A}}^{*}$ \citep{UlichHaas1976}. In the K-band, the calibration uncertainty is generally $\sim30\%$ for the GBT \citep{McGuire2020, Sita2022}.

\begin{figure*}
\begin{minipage}{9cm}
\includegraphics[width=\hsize]{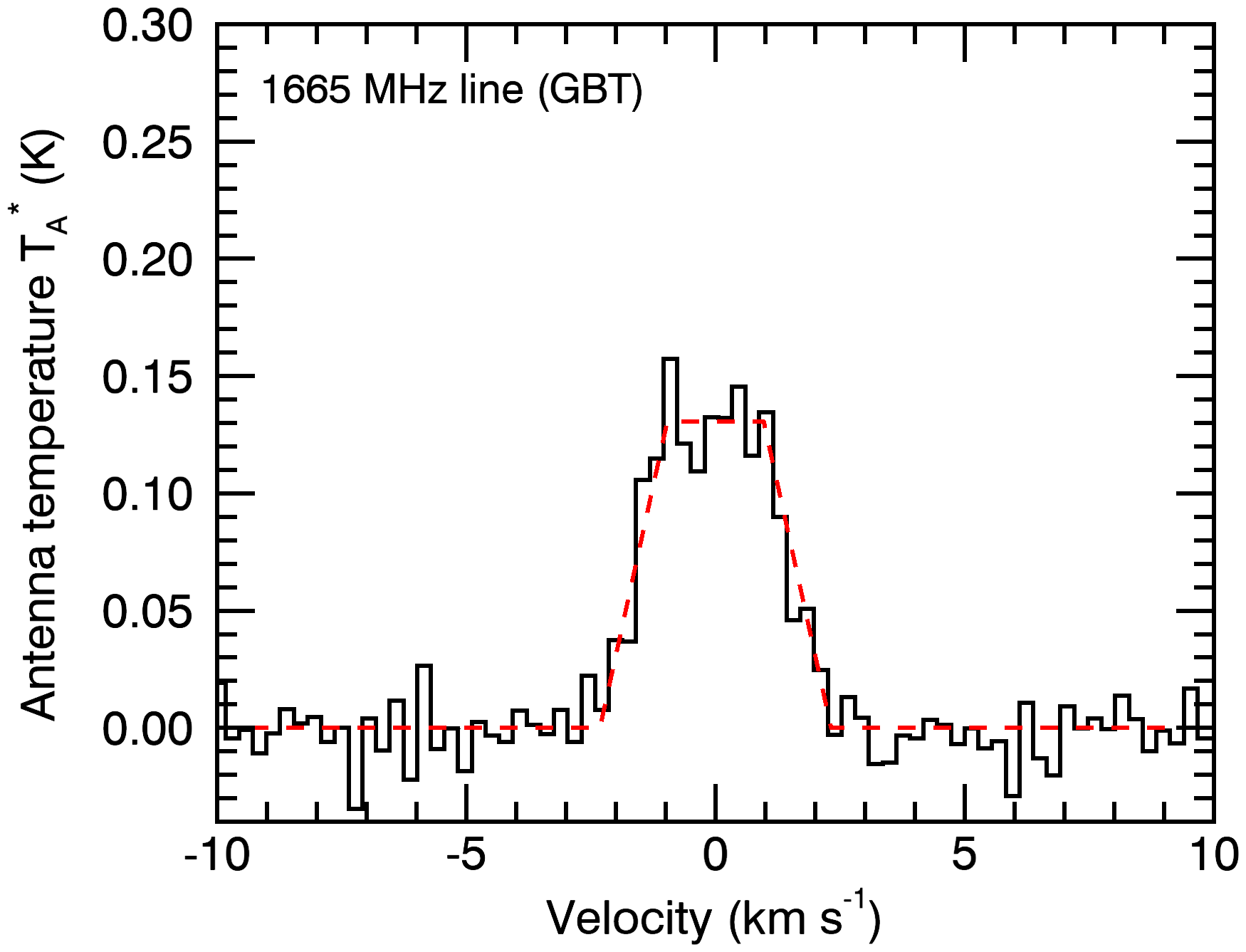}
\end{minipage}\hfill
\begin{minipage}{9cm}
\includegraphics[width=\hsize]{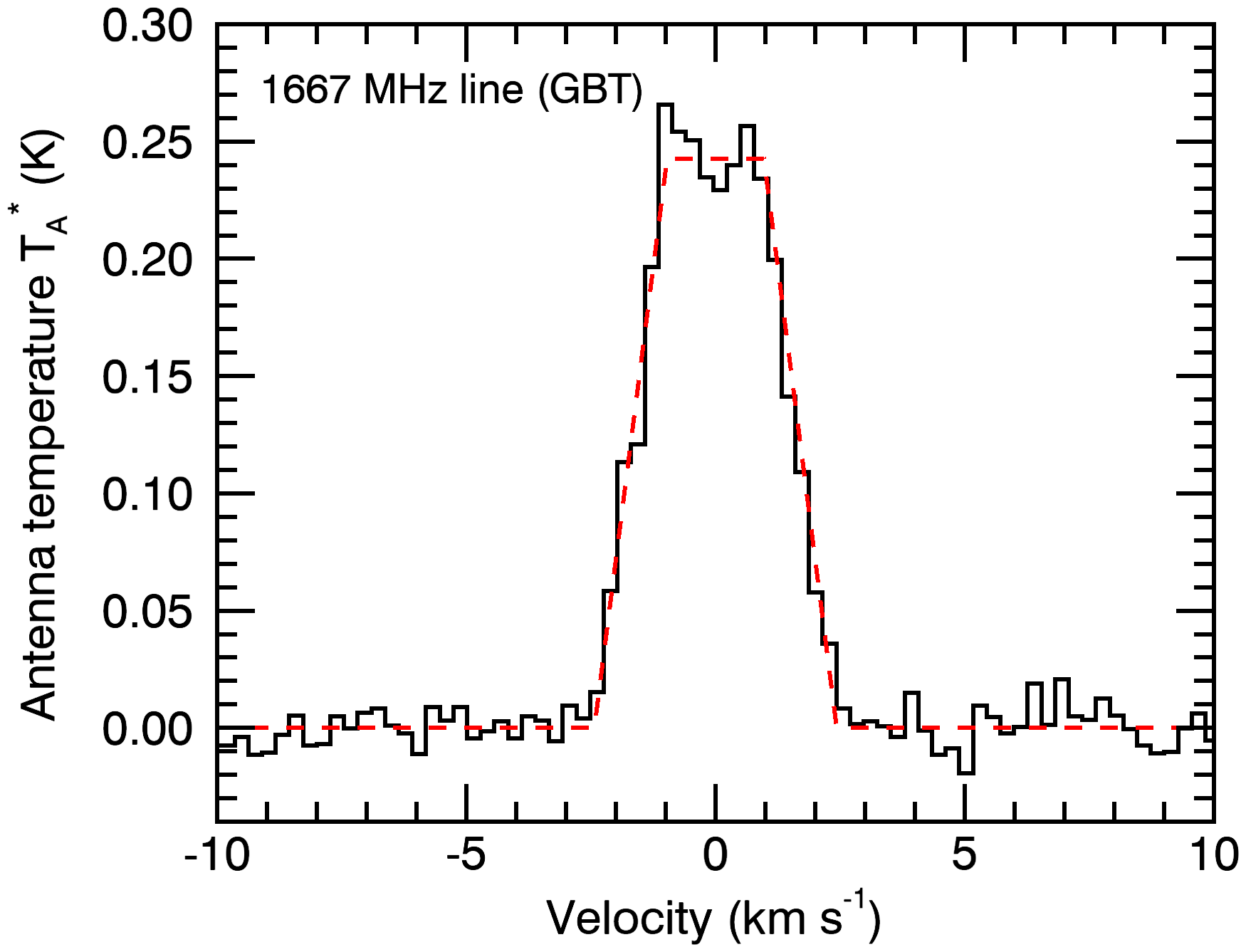}
\end{minipage}\hfill
\begin{minipage}{9cm}
\includegraphics[width=\hsize]{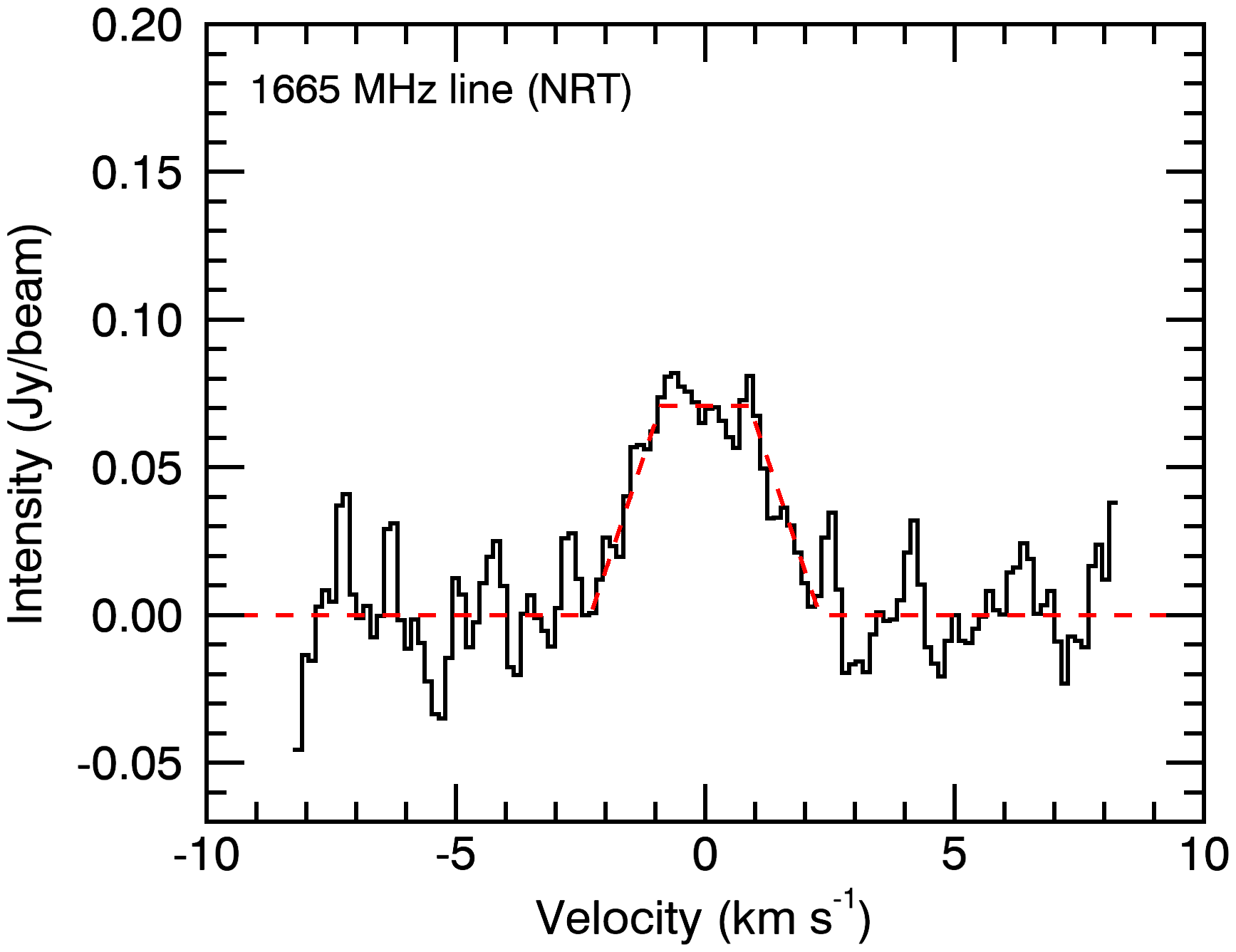}
\end{minipage}\hfill
\begin{minipage}{9cm}
\includegraphics[width=\hsize]{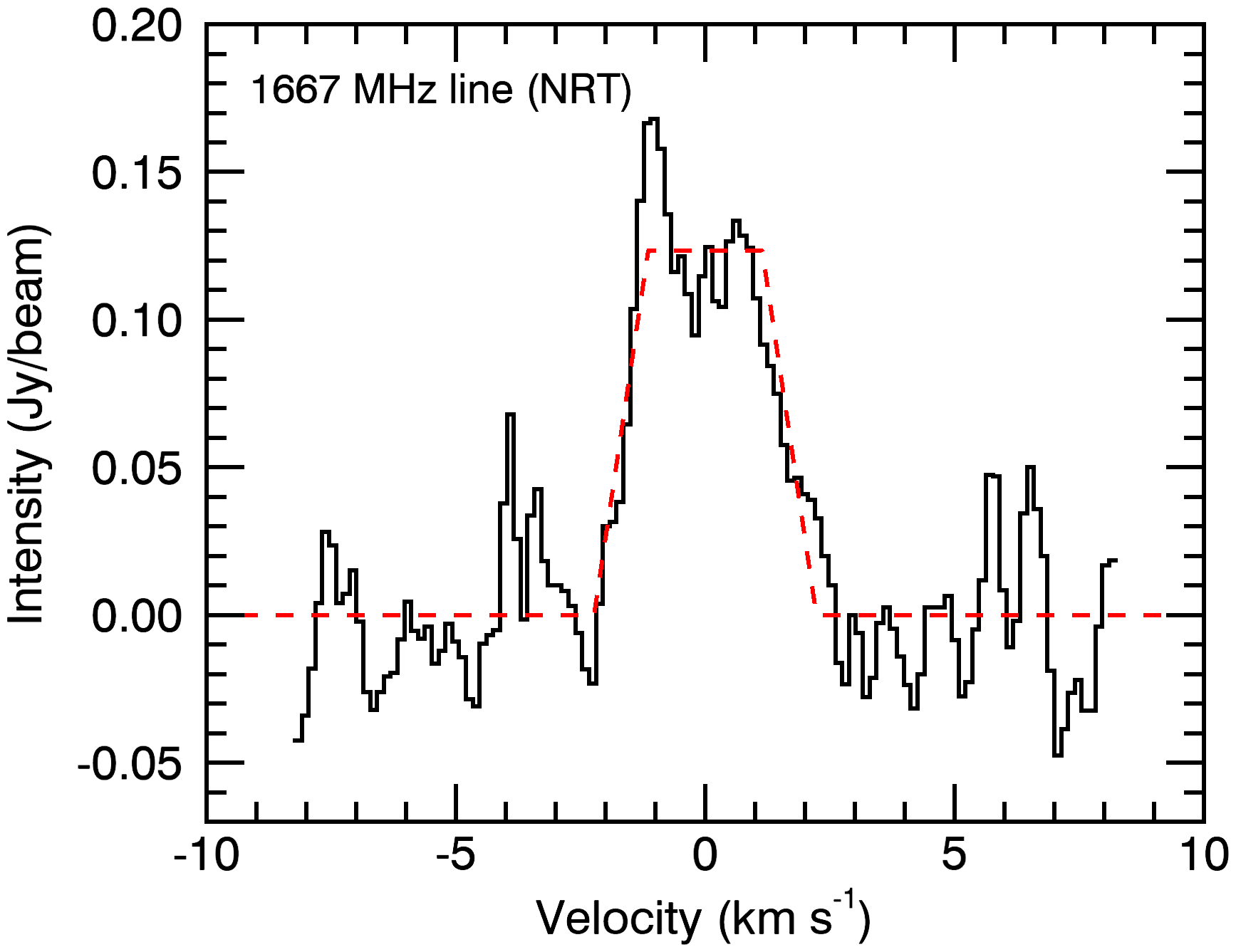}
\end{minipage}\hfill
\caption{OH $1~665$ (left) and $1~667$~MHz (right) lines observed with the GBT (top) and the NRT (bottom) on 24 July 2020 in solid black lines. The red dashed curves show the fitted trapeziums to the spectra based on the methodology of \citet{Bockelee-Morvan1990}. For GBT (NRT), the half lower bases of the trapeziums are $2.32 \pm 0.07$ ($2.30 \pm 0.14$) and $2.43 \pm 0.03$ ($2.23 \pm 0.11$)~km~s$^{-1}$ for the $1~665$ and $1~667$~MHz lines, respectively.}
\label{fig:trapeziumfitting}
\end{figure*}

\begin{figure}
\includegraphics[width=\hsize]{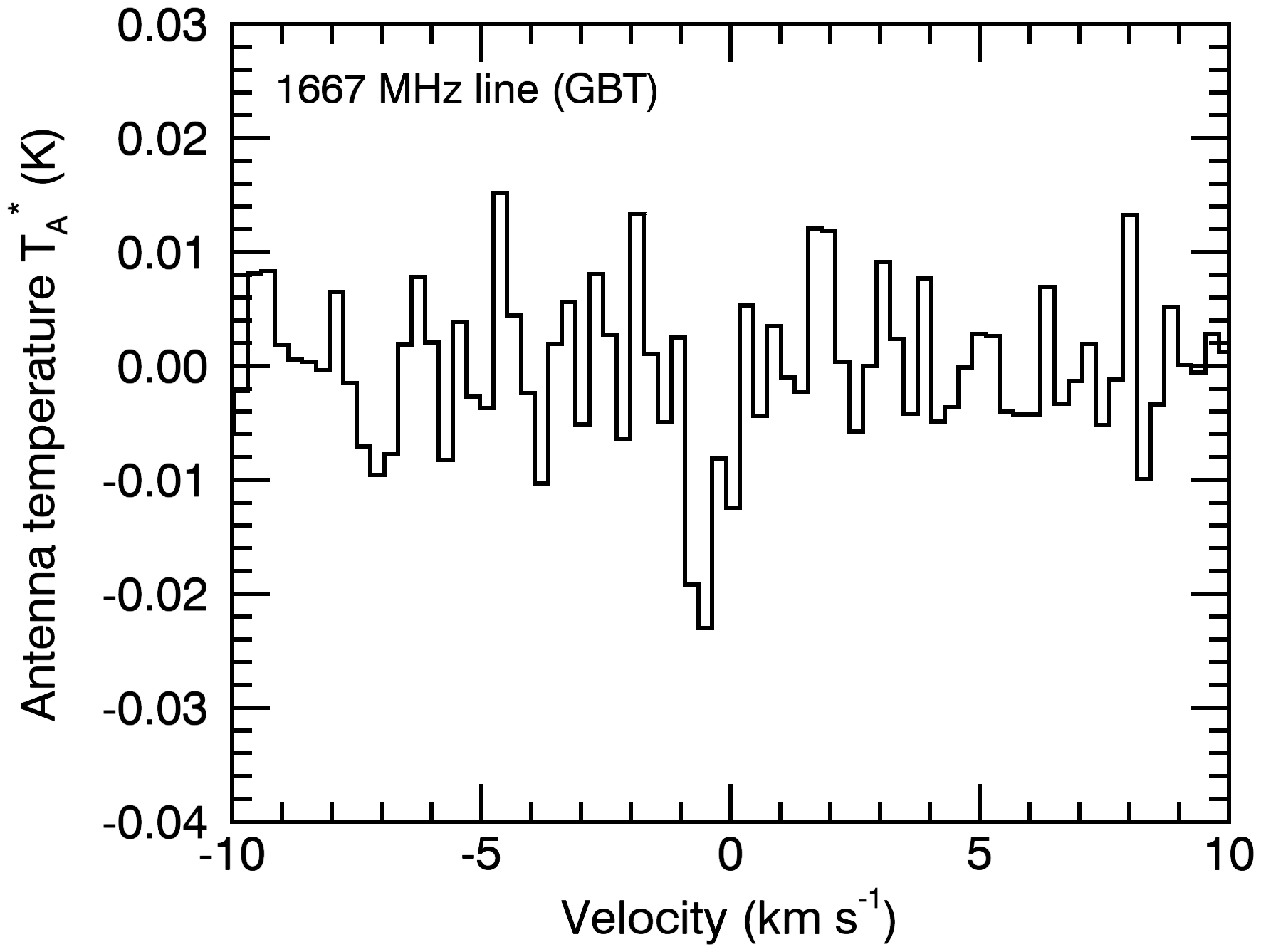}
\caption{OH $1~667$~MHz line observed with the GBT on 11.88 August 2020.}
\label{fig:OH-day2}
\end{figure}


\subsection{Nan\c{c}ay Radio Telescope observations}
\label{NRT}

\begin{figure*}
\centering
\includegraphics[width=0.78\hsize]{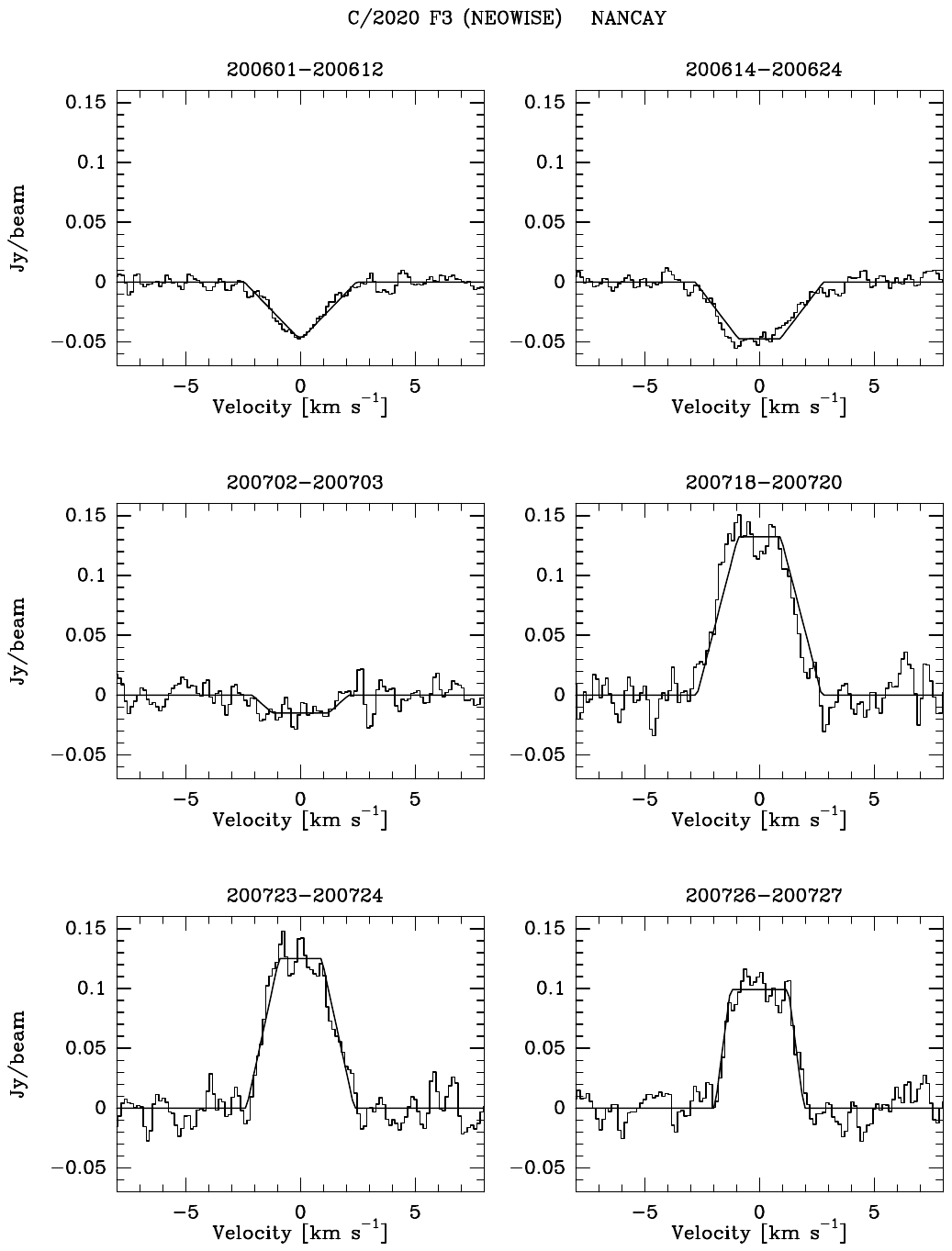}
\caption{Weighted averages of the OH $1~665$ and $1~6657$~MHz lines observed at the NRT, normalized to the intensity of the $1~667$~MHz (Section~\ref{NRT}). The spectra are integrated over specified periods of time, which are indicated above the plots in yymmdd format. The observations are fitted with trapeziums based on the methodology of \citet{Bockelee-Morvan1990}.}
\label{fig:OH-NRT}
\end{figure*}

Comet F3 was scheduled at the NRT as a Target of Opportunity (ToO) for observations beginning 1 June 2020 and continuing until 27 July 2020. It was observed almost every day during this period, except for 4-17 July 2020 (Table~\ref{table:lineareaNRT}). The instrumental characteristics, observing protocol, and data reduction procedure are the same as those used in preceding cometary observations with the NRT. These are described in \citet{Crovisier2002a} and \citet{Crovisier2002b}. The NRT is a Meridian telescope, which can observe a given source for $\sim 1$~h. Its RA$\times$Dec beam size is $3.5$\arcmin$ \times 19$\arcmin and its sensitivity is $0.85$~K~Jy$^{-1}$ at the 18-cm wavelength. The spectrometer, which can accommodate $8$ banks with each having a $195$~kHz bandwidth and $1~024$ channels, was aimed at the four OH transitions at $1~665$ and $1~667$~MHz (main lines) and $1~612$ and $1~721$~MHz (satellite lines) in, both, left- and right-hand circular polarizations with a $0.4$~kHz ($0.07$~km~s$^{-1}$ at $1~666.0$~MHz) spectral resolution after Hanning smoothing.

Fig.~\ref{fig:OH-NRT} shows the NRT spectra averaged over a few days that were used for kinematic studies based on the line shapes. The depicted spectra are averages of both polarizations and weighted averages of the $1~665$ and $1~667$~MHz main lines, converted to the $1~667$~MHz intensity scale by assuming the statistical ratio of $1.8$. The satellite $1~612$~MHz line is clearly detected in the NRT data averaged over the 18-27 July observing window. The spectrum containing the $1~720$~MHz satellite line is noisier; and the detection is only marginal. The four OH $18$-cm lines observed in comet F3 with the NRT averaged from 18 to 27 July 2020 are shown in Fig.~\ref{fig:aveNRT}. The expected statistical relative intensities of the $1~667:1~665:1~612:1~720$~MHz lines are $9:5:1:1$, which is in good agreement with the observations. The satellite OH lines were tentatively detected at the $1\sigma$ level on 31.8 July 2020 with the $\sim3\arcmin$ circular beam (at $1~667$~MHz) of the Arecibo Telescope \citep{Smith2021}.

The NRT spectra of 24.61 July 2020 were recorded nearly simultaneously with the Day 1 GBT observation and are analyzed individually in Section~\ref{sec:combined-OH-analysis} (sixth line of Table~\ref{table:lineareaNRT}). The $1~665$ and $1~667$~MHz spectra observed on this date are shown in Fig.~\ref{fig:trapeziumfitting}. The measured $1~667/1~665$~MHz line integrated intensity ratio is $1.82 \pm 0.13$, which is consistent within uncertainties with the statistical ratio of $1.8$. The other daily spectra are not shown individually. However, in Table~\ref{table:lineareaNRT}, daily measurements for July are provided, alongside averages over a few days in June and July. For the dates in June, it was not possible to compute the OH production rate on a daily basis due to the
relative weakness of the OH lines (due to a larger geocentric distance). Of particular importance is the entry for 2 July 2020 with its high maser inversion ($i<-0.31$), which allowed the OH production rate near perihelion to be estimated (Section~\ref{sec:NRT-OHprod}).


\section{Results}
\label{results}
\subsection{OH data analysis}
\label{OH}
\subsubsection{Model for OH production rate determination}
\label{OHprodrate}

In comets, the excitation of OH through UV pumping and subsequent fluorescence leads to population inversion (or anti-inversion) in the sublevels of the $\Lambda$-doublet of the $^2\Pi_{3/2}~J=3/2$ ground-state. The population inversion ($i$) depends strongly on the comet heliocentric velocity and has been modeled by \citet{Despois1981} and \citet{SchleicherAHearn1988}. Depending on the sign of $i$, the four hyperfine components of the $\Lambda$-doublet appear in either emission or absorption. The velocity-integrated flux density (i.e., line area) can be expressed as \citep[e.g.,][]{Despois1981, SchloerbGerard1985}:
\begin{equation}
\int S_\nu dv= \frac{A_{\rm ul}  k c T_{\rm bg}}{4 \pi \Delta^2} \frac{2F_u+1}{8}  \Big[i+\frac{i+1}{2}\frac{h\nu}{k T_{\rm bg}}\Big]~\Gamma_{\rm OH}.
\label{eq:1}
\end{equation}
\noindent
In this equation, $A_{\rm ul}$ is the Einstein spontaneous coefficient of the line at frequency $\nu$, $F_{\rm u}$ is the statistical weight of the upper level of the transition ($F_{\rm u} = 1$ and $2$ for the $1~667$ and $1~665$~MHz lines, respectively), $i$ is the inversion of the ground state $\Lambda$-doublet levels, $T_{\rm bg}$ is the background temperature, and $\Delta$ is the Earth-comet distance. $\Gamma_{\rm OH}$ is the number of molecules within the beam, which is proportional to the OH production rate ($Q_{\rm OH}$). The second term inside the brackets corresponds to spontaneous emission and is only significant for small values of $i$. For the 100-m GBT, the area of the $1~667$~MHz line, expressed in K~km~s$^{-1}$ on the main beam brightness temperature scale, is given by:
\begin{equation}
\int T_{\rm mB} dv=  9.157 \times 10^{-35} \frac{T_{\rm bg}}{{\Delta^2}} \Big[i+\frac{i+1}{2}\frac{h\nu}{k T_{\rm bg}}\Big]~\Gamma_{\rm OH}, 
\label{eq:2}
\end{equation}
\noindent
where the geocentric distance is in astronomical units.

In the inner part of the coma, collisions thermalize the populations of the $\Lambda$-doublet and quench the signal. Based on 1P/Halley observations, \citet{Gerard1990} showed that the quenching radius (in km) can be approximated by:
\begin{equation}
 r_{\rm q} = 4.7 \times 10^{4} r_{\rm h} \sqrt{Q_{\rm OH}/10^{29}}, 
 \label{eq:quenching}
 \end{equation} 
\noindent
where $Q_{\rm OH}$ is the OH production rate in molec~s$^{-1}$. For an active comet such as F3, the quenched region can be expected to be comparable to the projected field of view, which is $2.29 \times 10^{5}$~km in diameter for the GBT and $\left( 1.07 \times 10^{5} \right) \times \left( 5.80 \times 10^{5} \right)$~km for the NRT on 24 July 2020. Hence, collisional quenching is an important factor to consider when deriving OH production rates from the GBT and NRT observations of comet F3. The nearly simultaneous observations (a mere $2$~hours apart) of the comet by these two instruments provide a new opportunity to measure the quenching radius. The very few prior measurements are based on the comparison of 18-cm- and UV-derived OH production rates \citep{Gerard1990} and on the analysis of 18-cm observations taken at various offset positions from the comet nucleus \citep{Colom1999, Schloerb1997, Gerard1998}. 

In order to take into account collisional quenching, it is assumed that the maser inversion is zero for cometocentric distances $r \leq r_{\rm q}$, following \citet{Schloerb1988} and \citet{Gerard1990}. More realistic descriptions with a progressive quenching throughout the coma have been investigated by \citet{Colom1999} and \citet{Gerard1998}. Equations~\ref{eq:1}-\ref{eq:2} are then no longer valid, but can be replaced by similar equations: $\Gamma_{\rm OH}$ is replaced by the number of unquenched OH radicals within the beam and only the spontaneous emission term (with $i=0$) is considered for the OH radicals within the quenched region. \citet{Despois1981} have shown that in the collisional region the maser inversion is very small, that is on the order of $10^{-4}$ for a kinetic temperature of $300$~K; thus, it can be safely neglected for the purpose of our analysis of comet F3. The calculation of the number of quenched and unquenched OH radicals within the beam is done by volume integration within a Gaussian beam using a description of the distribution of the OH radicals in the coma \citep[see, e.g.,][]{Despois1981}.

For the spatial distribution of the OH radicals, the Haser-equivalent model is employed \citep{CombiDelsemme1980a}. The water and OH radical lifetimes at $1$~au from the Sun are set to $8.5 \times 10^4$ (for the quiet Sun from \citealt{Crovisier1989}, considering that the Sun was quiet in summer 2020) and $1.1 \times 10^5$~s (\citealt{vanDishoeckDalgarno1984}, but poorly constrained as discussed in \citealt{SchloerbGerard1985}), respectively (see also table~$3$ in \citealt{Crovisier2002a})\footnote{Using the updated value from \citet{Heays2017}, OH radical lifetime at $1$~au from the Sun is $1.6 \times 10^5$~s.}. The OH ejection velocity is set to $v_{\rm e} = 0.95$~km~s$^{-1}$ \citep{Bockelee-Morvan1990, Crovisier2002a}. The OH parent velocity $v_{\rm p}$ is derived from the OH line profile, as described in Section~\ref{sec:lineprofile}.

\begin{table*}
\caption{GBT observations: line intensities and spectral characteristics.}
\label{table:linearea}
\begin{tabular}{lcccccccc}
\hline\hline
UT Date & $\Delta$ & $r_{\text{h}}$ & Line  & RMS$^{(1)}$     & Area ($T_{\text{A}}^{*})^{(2)}$ & Width$^{(3)}$ & $\Delta v^{(3)}$ & $v_{p}$+$v_{e}^{(4)}$\\
(yyyy/mm/dd)     & (au)       & (au)             &       & (mK channel$^{-1}$)  & (K km s$^{-1}$)                   & (km s$^{-1}$)   & (km s$^{-1}$)      & (km s$^{-1}$)\\ 
\hline\noalign{\vskip 2mm}
2020$/$07$/$24.71 & 0.696  & 0.667 & OH $1~665$ & $12$ & $0.428 \pm 0.014$ & $2.89 \pm 0.11$ & $-0.07 \pm 0.05$ & $2.32 \pm 0.07$\\
2020$/$08$/$11.88 & 1.078  & 1.053 & OH $1~665$ & $9$  & $<0.03$ & -- & -- & --\\
2020$/$07$/$24.71 & 0.696  & 0.667 & OH $1~667$ & $10$ & $0.827 \pm 0.012$ & $2.95 \pm 0.05$ & $-0.05 \pm 0.02$ & $2.43 \pm 0.04$\\
2020$/$08$/$11.88 & 1.078  & 1.053 & OH $1~667$ & $9$  & $-0.017 \pm 0.005$ & $0.67 \pm 0.27$ & $-0.53 \pm 0.12$ & --\\
2020$/$07$/$24.81 & 0.696  & 0.669 & NH$_3$ ($1_1-1_1$) & $48$ & $<0.017$ & -- & -- & --\\
2020$/$07$/$24.81 & 0.696  & 0.669 & NH$_3$ ($2_2-2_2$) & $47$ & $<0.016$ & -- & -- & --\\
2020$/$07$/$24.81 & 0.696  & 0.669 & NH$_3$ ($3_3-3_3$) & $47$ & $<0.016$ & -- & -- & --\\
2020$/$07$/$24.81 & 0.696  & 0.669 & NH$_3$ ($4_4-4_4$) & $46$ & $<0.016$ & -- & -- & --\\
2020$/$07$/$24.81 & 0.696  & 0.669 & NH$_3$ ($5_5-5_5$) & $46$ & $<0.016$ & -- & -- & --\\
\hline
\end{tabular}

\footnotesize{
$^{(1)}$ For a spectral channel width of $\sim0.005$~km~s$^{-1}$ for the NH$_3$ lines and $\sim0.3$~km~s$^{-1}$ for the OH $1~665$ and $1~667$~MHz lines.\\
$^{(2)}$ Velocity-integrated flux density (i.e., line area) in the $T_{\text{A}}^{*}$ scale or a $3\sigma$ upper limit. Considered velocity intervals are $\left[ -2.5, +2.5 \right]$~km~s$^{-1}$ and $\left[-1.5, 1.5 \right]$~km~s$^{-1}$ for the OH and NH$_{3}$ lines, respectively. For the OH $1~667$~MHz line observed on 11.88 August 2020, the line area is that obtained from a Gaussian fit.\\
$^{(3)}$ Velocity offset from a Gaussian fit.\\
$^{(4)}$ Half lower bases of the fitted trapezia (Fig.~\ref{fig:trapeziumfitting}).}
\end{table*}

\begin{sidewaystable*}
\centering
\caption{OH observations at NRT: line intensities and spectral characteristics.}
\label{table:lineareaNRT}
\begin{tabular}{lccrrrccccll}
\hline\hline
UT Date  & $\Delta$ & $r_{\text{h}}$ & $v_{\text{h}}$ & $i^{(1)}$ & $i^{(2)}$ & $T_{\text{bg}}$ & Area$^{(3)}$    & $\Delta v^{(4)}$ & $v_{p}+v_{e}^{(5)}$ & $Q_{\text{OH}}^{(1)}$ & $Q_{\text{OH}}^{(2)}$\B\\
(yyyy/mm/dd) & (au)       & (au)             & (km~s$^{-1}$)    &           &           & (K)               & (mJy km s$^{-1}$) & (km~s$^{-1}$)      & (km s$^{-1}$)         & ($10^{29}$ molec~s$^{-1}$)    & ($10^{29}$ molec~s$^{-1}$)\\
&& && (Des) & (Sch) &&&&& (Des) & (Sch)\\
\hline\noalign{\vskip 2mm}
2020/07/02.46 & 1.19 & 0.30 & $-7.2$ & $-0.31$ & $-0.37$ & 3.4 & $-30 \pm 13$ & --               & --              & $18.6 \pm 4.8$        & $22.0 \pm 4.8$\\
2020/07/18.54 & 0.72 & 0.53 & 38.5   & 0.20    & 0.28    & 3.1 & $526 \pm 32$ & $-0.41 \pm 0.14$ & --              & $21.4 \pm 3.8^{(6)}$  & $8.47 \pm 1.17^{(6)}$\\
2020/07/19.55 & 0.71 & 0.55 & 38.7   & 0.22    & 0.30    & 3.1 & $379 \pm 32$ & $-0.13 \pm 0.16$ & --              & $6.03 \pm 0.93^{(6)}$ & $3.69 \pm 0.48^{(6)}$\\
2020/07/20.56 & 0.70 & 0.57 & 38.8   & 0.23    & 0.30    & 3.1 & $491 \pm 35$ & $-0.33 \pm 0.13$ & --              & $8.81 \pm 1.62^{(6)}$ & $5.14 \pm 0.69^{(6)}$\\
2020/07/23.60 & 0.69 & 0.64 & 38.6   & 0.21    & 0.29    & 3.1 & $402 \pm 20$ & $-0.07 \pm 0.09$ & --              & $5.13 \pm 0.53^{(6)}$ & $3.00 \pm 0.23^{(6)}$\\
2020/07/24.61 & 0.70 & 0.66 & 38.5   & 0.20    & 0.28    & 3.1 & $414 \pm 23$ & $-0.12 \pm 0.09$ & $2.25 \pm 0.14$ & $4.62 \pm 0.38$       & $3.41 \pm 0.25$\\
2020/07/26.63 & 0.71 & 0.71 & 38.2   & 0.17    & 0.25    & 3.1 & $359 \pm 19$ & $+0.05 \pm 0.10$ & --              & $5.15 \pm 0.59^{(6)}$ & $2.71 \pm 0.23^{(6)}$\\
2020/07/27.63 & 0.72 & 0.73 & 38.0   & 0.15    & 0.23    & 3.1 & $279 \pm 19$ & $-0.10 \pm 0.11$ & --              & $3.79 \pm 0.47^{(6)}$ & $2.02 \pm 0.20^{(6)}$\\
\hline\noalign{\vskip 2mm}  
2020/06/01-2020/06/12 & 1.58 & 0.79 & $-37.5$ & $-0.24$ & $-0.26$ & 3.4 & $-115 \pm 5$ & $-0.09 \pm 0.06$ & $2.44 \pm 0.08$ & $2.32 \pm 0.12$ & $2.10 \pm 0.11$\\
2020/06/14-2020/06/24 & 1.47 & 0.52 & $-37.7$ & $-0.25$ & $-0.26$ & 3.5 & $-165 \pm 5$ & $-0.20 \pm 0.07$ & $2.79 \pm 0.11$ & $8.00 \pm 0.49$ & $7.32 \pm 0.50$\\
2020/07/02-2020/07/03 & 1.18 & 0.30 & $-4.8$ & $-0.23$ & $-0.28$ & 3.4 & $-49 \pm 10$ & -- & -- & -- & --\\
2020/07/18-2020/07/20 & 0.71 & 0.55 & 38.6 & 0.21 & 0.29 & 3.1 & $459 \pm 19$ & $-0.28 \pm 0.08$ & $2.72 \pm 0.12$ & $10.0 \pm 1.1$ & $5.30 \pm 0.37$\\
2020/07/23-2020/07/24 & 0.69 & 0.65 & 38.6 & 0.21 & 0.29 & 3.1 & $409 \pm 14$ & $-0.09 \pm 0.06$ & $2.35 \pm 0.10$ & $5.03 \pm 0.34$ & $3.03 \pm 0.17$\\
2020/07/26-2020/07/27 & 0.72 & 0.72 & 38.1 & 0.16 & 0.24 & 3.1 & $322 \pm 13$ & $-0.02 \pm 0.08$ & $1.94 \pm 0.05$ & $4.62 \pm 0.41$ & $2.40 \pm 0.15$\\
\hline
\end{tabular}
\footnotesize{\\
$^{(1)}$ Maser inversion from \citet{Despois1981} and corresponding inferred OH production rate.\\
$^{(2)}$ Maser inversion from \citet{SchleicherAHearn1988} and corresponding inferred OH production rate.\\
$^{(3)}$ Weighted average of the $1~667$ and $1~665$~MHz lines converted to the $1~667$~MHz intensity scale, assuming the statistical ratio of $1.8$.\\
$^{(4)}$ Velocity offset from a Gaussian fit.\\
$^{(5)}$ Half lower bases of the fitted trapezia (Figs.~\ref{fig:trapeziumfitting} and~\ref{fig:OH-NRT}).\\
$^{(6)}$ Trapezium fitting was not performed for this date. The assumed $v_{p}$ value is that deduced from the $r_{\text{h}}$ evolution of $v_{p}$ measured from the $2-3$ day averages over the $18-27$ July 2020 period.}
\end{sidewaystable*}


\subsubsection{Line profiles and H$_{2}$O outflow velocity}
\label{sec:lineprofile}

The observed line shapes have been analyzed in the framework of the trapezium modeling as proposed by \citet{Bockelee-Morvan1990} and subsequently applied to the kinematic studies of the coma of many comets in \citet{Tseng2007}. OH is a daughter species of H$_{2}$O photodissociation and is assumed here to be emitted isotropically in the rest frame of H$_{2}$O. Indeed, some anisotropy may occur since water photodissociation is caused by the unidirectional solar UV radiation (e.g., \citealt{Crovisier1990}). However, the currently available theoretical work and laboratory data do not permit a complete evaluation of this potential anisotropy. Taking into account collisional quenching, the line width of OH does not provide the velocity of its parent molecule. The maximum radial velocity of OH along the line of sight is $v_{\rm p}$+$v_{\rm e}$, assuming that the OH parent and OH ejection velocity distributions are monokinetic. A trapezium centered on the cometary nucleus is expected when the beam is very large with respect to the OH coma. As shown by \citet{Bockelee-Morvan1990}, the half lower base of the fitted trapezium to an OH line is expected to be equal to $v_{\rm p}$+$v_{\rm e}$. Figs.~\ref{fig:trapeziumfitting} and \ref{fig:OH-NRT} show the trapezium method applied to the GBT and NRT OH spectra. The derived $v_{\rm p}$+$v_{\rm e}$ values are given in Tables~\ref{table:linearea} and \ref{table:lineareaNRT}.

The OH parent velocity derived from the GBT $1~667$~MHz line observed on 24 July 2020 is $v_{\rm p} = 1.48 \pm 0.04$~km~s$^{-1}$. The trapezium method applied to the Nan\c{c}ay spectrum of 24 July 2020 yields $v_{\rm p}$ = 1.30 $\pm$ 0.14 km s$^{-1}$, which is consistent within $1\sigma$ with the GBT-derived value. A slightly lower $v_{\rm p}$ value measured at NRT is not unexpected, as it could be explained by gas acceleration in the coma. With its $3.5\arcmin \times 19\arcmin$ beam, the NRT field of view is probing OH radicals closer to the surface than the $7.55\arcmin$ GBT beam (at $1~667$~MHz). As expected, the OH parent expansion velocity $v_{\rm p}$ is observed to increase from $\sim 1.0$ to $\sim 1.8$~km~s$^{-1}$ when the heliocentric distance $r_{\rm h}$ decreases from $\sim 0.72$ to $\sim 0.52$~au (Table~\ref{table:lineareaNRT}). This trend was observed in other comets \citep{Tseng2007}.


\subsubsection{Combined analysis of the GBT and NRT OH observations of 24 July 2020}
\label{sec:combined-OH-analysis}

\begin{figure}
\includegraphics[width=\hsize]{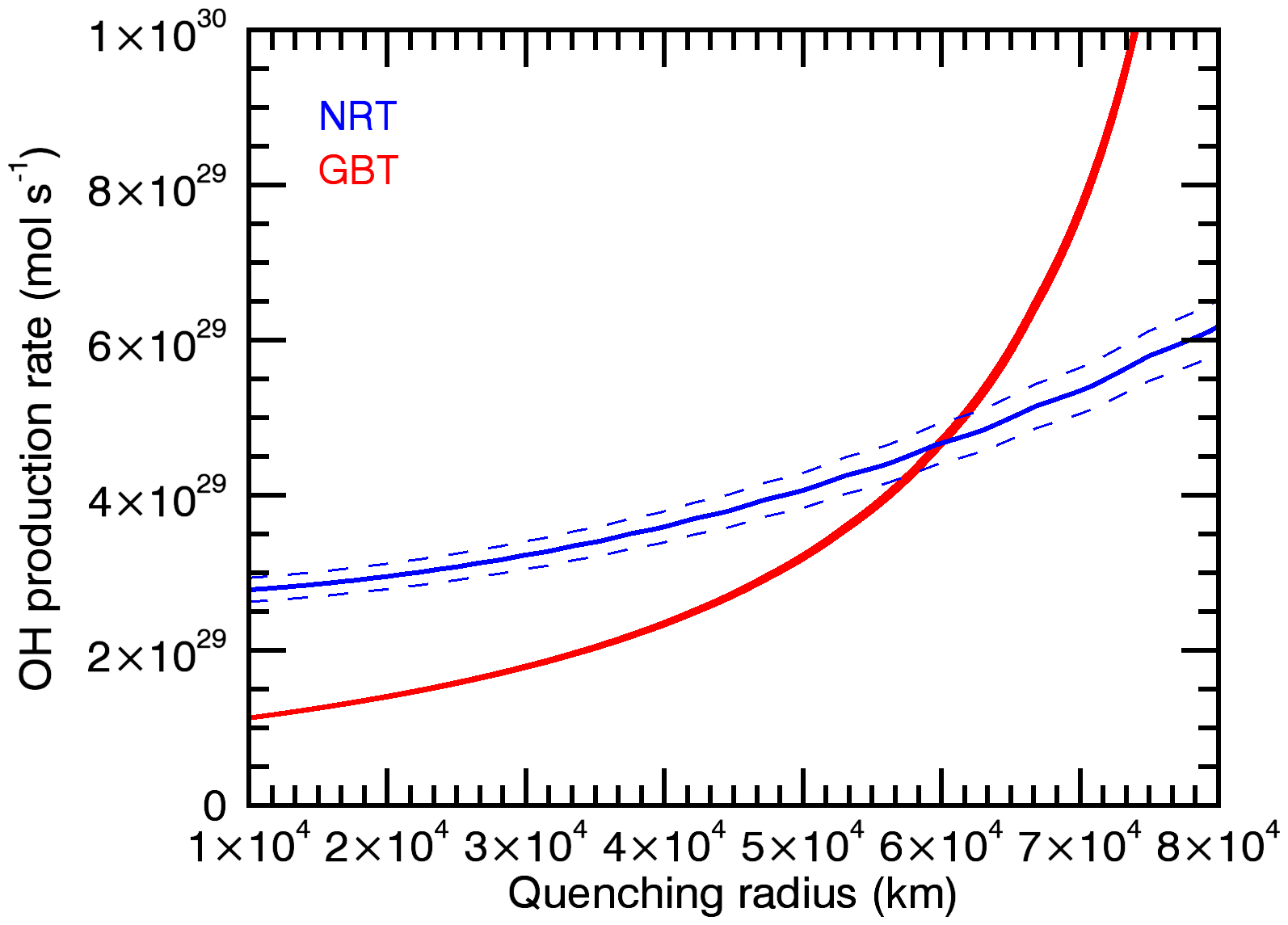}
\caption{Derived OH production rates from the GBT (red curve) and NRT (blue curves) observations of the OH $1~667$~MHz line in comet C/2020 F3 (NEOWISE) on 24 July 2020. OH production rates are plotted as a function of the quenching radius. The dashed lines correspond to the $\pm 1\sigma$ uncertainty in the line area measured at NRT. Analogous dashed lines for the GBT are indistinguishable from the solid curve due to the high signal-to-noise ratio of the data. The maser inversion is $i = 0.20$ based on \citet{Despois1981}, and the background temperature is $T_{\text{bg}} = 3.1$~K. The OH spatial distribution is described by the Haser-equivalent model \citep{CombiDelsemme1980a}.}
\label{fig:rq-QOH}
\end{figure}

Due to the significant differences in beam sizes and shapes between the GBT and NRT (of a factor of about two considering the small dimension of the NRT's elliptical beam), the fractions of quenched OH radicals that the two telescopes sample differ. The quenching radius $r_{\rm q}$ can be determined by searching for the $Q_{\rm OH}$ value that is consistent with data from both facilities. Fig.~\ref{fig:rq-QOH} shows $Q_{\rm OH}$ values derived as a function of $r_{\rm q}$ from the 24 July 2020 GBT and NRT data using the maser inversion of $i = 0.20$ based on \citet{Despois1981}. For the GBT and NRT calculations, the derived $v_{\rm p} = 1.48$~km~s$^{-1}$ and $v_{\rm p} = 1.30$~km s$^{-1}$ values are utilized, respectively (Section~\ref{sec:lineprofile}). The obtained quenching radius is $\left( 5.95 \pm 0.14 \right) \times 10^{4}$~km and the consistent OH production rate is $\left( 4.62 \pm 0.38 \right) \times 10^{29}$~molec~s$^{-1}$. Using the maser inversion $i = 0.28$ based on \citet{SchleicherAHearn1988}, the derived $r_{\rm q}$ value is the same ($\left( 5.96 \pm 0.14 \right) \times 10^{4}$~km), but the OH production rate is $\left( 3.41 \pm 0.25 \right) \times 10^{29}$~molec~s$^{-1}$. The derived quenching radius is consistent with the law provided by Eq.~\ref{eq:quenching}, which predicts values of $5.6 \times 10^{4}$ and $6.6 \times 10^{4}$~km for the two obtained $Q_{\rm OH}$ values. Averaging the production rates inferred with the \citet{Despois1981} and \citet{SchleicherAHearn1988} maser inversion ($i$) models yields an OH production rate of $\left(4.02\pm0.32\right)\times10^{29}$~molec~s$^{-1}$, which is consistent with each model individually within errors. The corresponding averaged water production rate is $\left(4.42\pm0.35\right)\times10^{29}$~molec~s$^{-1}$. The two considered maser inversion models were developed independently and compared in \citet{SchloerbGerard1985}. The two models use different sources for the solar spectrum. The model of \citet{SchleicherAHearn1988} includes IR pumping (a minor effect). Both models give remarkably similar results, except when the OH inversion is low.


\subsubsection{GBT spectrum of 11 August 2020}
\label{GBTAug11}

The GBT on-nucleus data obtained on 11 August 2020 show a marginal absorption line near the frequency of the OH $1~667$~MHz line at the nucleus-centered beam position (Fig.~\ref{fig:OH-day2}). Unlike the 24 July 2020 spectrum, the line is narrow ($\sim1$~km~s$^{-1}$), strongly blueshifted ($\Delta v =-0.53$~km~s$^{-1}$ from a Gaussian fit, Table~\ref{table:GBT}), and does not show any signal at positive Doppler velocities. This line shape can be explained by the Greenstein effect \citep{Greenstein1958}. As a result of the motion of the OH radicals in the coma, their heliocentric velocity is shifted with respect to the nucleus heliocentric velocity, adding another component to the maser inversion \citep{Despois1981}. In most instances, the Greenstein effect only affects weakly the shape of the OH lines. However, the effect is striking at heliocentric velocities where the maser inversion changes its sign over a small velocity range, since some radicals have a positive maser inversion, and others have a negative inversion.

The heliocentric velocity of comet F3 at the time when the GBT observations of 11 August 2020 were carried out was $v_{\text{h}} = +34.8$~km~s$^{-1}$. For the \citet{Despois1981} model, the maser inversion is $i=-0.125$ at this $v_{\text{h}}$, but varies from $-0.240$ to $+0.006$ in the coma assuming that the maximum OH expansion velocity ($v_{\rm p}$+$v_{\rm e}$) is $2$~km~s$^{-1}$. For the \citet{SchleicherAHearn1988} model, the inversion is in the $\left[-0.16, +0.09\right]$ range with $i=-0.020$ at this value of $v_{\text{h}}$. Taking into account the Sun-comet-Earth angle of $58.8^{\circ}$ (phase angle) and spontaneous emission, it is expected that the line will be in absorption at negative Doppler velocities and will show at positive velocities either weak positive or negative emission depending on the Schleicher et al. or the Despois et al. inversion values, respectively. The low signal-to-noise ratio prevents any conclusion about which model is in better accordance with the observed spectrum.

Equations \ref{eq:1}-\ref{eq:2} hold in the presence of the Greenstein effect, so they have been used for an attempt to determine the OH production rate on 11 August 2020. For this, the quenching law given by Eq.~\ref{eq:quenching}, $T_{\rm bg} = 3.1$~K, and $v_{\rm p} = 1.2$~km~s$^{-1}$ have been employed. This $v_{\rm p}$ value is calculated from the value of  $1.48$~km~s$^{-1}$ determined on 24 July (Table~\ref{table:linearea}) and an assumed a $r_{\text{h}}^{-0.5}$ dependence. The Schleicher et al. excitation model cannot explain the observed line intensity (Table~\ref{table:linearea}) with this quenching law. A value of $Q_{\rm OH}$ in the $0.5-3 \times 10^{29}$~molec~s$^{-1}$ range is obtained using the Despois et al. inversion value. If the quenching radius is left as a free parameter, the derived $Q_{\rm OH}$ strongly depends on $r_{\rm q}$ and $i$ values. It becomes necessary to conclude that it is not possible to derive a reliable $Q_{\rm OH}$ value from the 11 August 2020 data. The $1~665$~MHz spectrum and the $1~665$, $1~667$~MHz spectra obtained at offset positions do not show even any marginal hints of lines and were not analyzed.


\subsubsection{Evolution of the OH production rate}
\label{sec:NRT-OHprod}

Table~\ref{table:lineareaNRT} presents the OH production rates determined from the NRT data using the \citet{Despois1981} and the \citet{SchleicherAHearn1988} inversion models. The average spectrum of 2-3 July 2020 observations shows an absorption line consistent with the inversion models. The maser inversion was very low on 3 July 2020. Consequently, as for the GBT data of 11 August 2020 (Section~\ref{GBTAug11}), this observation could not be used for determining a production rate. On the other hand, an OH production rate of $\sim2\times10^{30}$~molec~s$^{-1}$ could be determined for 2 July (i.e., one day before perihelion; Table~\ref{table:lineareaNRT}). Despite the maser inversion being large on this date (either $-0.31$ or $-0.37$ depending on the inversion model), the signal was weak due to a large fraction of quenched OH radicals in the beam.

The OH 18-cm lines were observed in emission with the Arecibo 305-m dish on 31.8 July 2020 ($r_{\rm h} = 0.82$~au), from which an unexpectedly low OH production rate of $\left( 3.6 \pm 0.6 \right) \times10^{28}$~molec~s$^{-1}$ was derived \citep{Smith2021}. However, collisional quenching was not taken into account in the analysis presented in that paper. In this work, the Arecibo data were reanalyzed with collision quenching being taken into account. The inferred values are $Q_{\rm OH} = \left(1.5\pm0.3\right)\times10^{29}$~molec~s$^{-1}$ with the inversion model of \citet{Despois1981} and $\left( 2.95\pm0.05 \right) \times10^{29}$~molec~s$^{-1}$ with that of \citet{SchleicherAHearn1988}. These values are in much closer agreement with the production rate estimated from optical OH line observations for the same observation date of $8.5 \times 10^{28}$~molec~s$^{-1}$ quoted in the \citet{Smith2021} paper (based on D. Schleicher 2021, personal communication).

\begin{figure}
\includegraphics[width=\hsize]{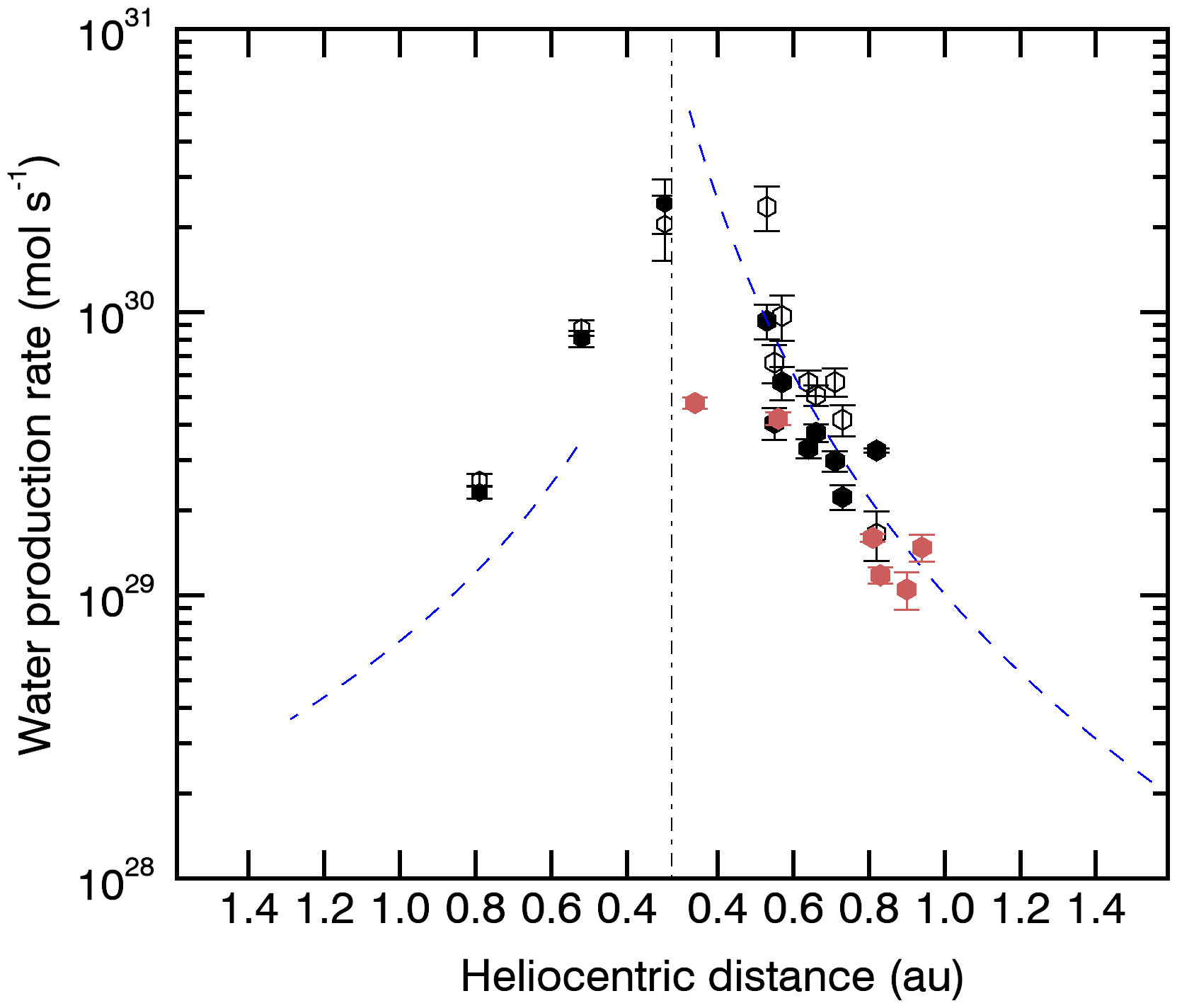}
\caption{Water production rates of comet C/2020 F3 (NEOWISE). Black dots are values from this work and include the revised value from the Arecibo data: filled and open  dots refer to values obtained with the inversion models of \citet{SchleicherAHearn1988} and \citet{Despois1981}, respectively. Red dots: values from near-IR observations from \citet{Faggi2021}. The dashed blue lines are a fit to the pre- and post-perihelion heliocentric values derived from Ly-$\alpha$ data \citep{Combi2021}.}
\label{fig:QH2O}
\end{figure}

The derived water production rates ($Q_{\text{H}_{2}\text{O}} = 1.1 \times Q_{\rm OH}$; \citealt{Crovisier1989}) are plotted in Fig.~\ref{fig:QH2O} together with other water production rate determinations from near-IR water lines observed with long-slit spectroscopy with the iSHELL at NASA/IRTF \citep{Faggi2021} and from Ly-$\alpha$ observations using SOHO/SWAN \citep{Combi2021}. There is a good agreement between the OH 18-cm and Ly-$\alpha$ $Q_{\text{H}_{2}\text{O}}$ data post-perihelion, but a factor of two discrepancy is observed pre-perihelion. Understanding this is potentially important, but requires dedicated modeling efforts that are beyond the scope of this paper. Near-IR determinations, which were all obtained post-perihelion (see red symbols in Fig.~\ref{fig:QH2O}), are overall consistent with the OH 18-cm for contemporaneous dates. The large discrepancy between near-IR and Ly-$\alpha$ measurements near-perihelion ($r_{\rm h}<0.4$~au) is discussed in \citet{Faggi2021} and may be related to an extended production of water, for example, from icy grains. Generally, robust comparisons across the UV, IR, and radio domains are challenging due to the significantly different spatial scales being probed (FOV of $30^{\circ}$, $0.3-4\arcsec$, and $32\arcsec-19\arcmin$, respectively).

\subsection{NH$_{3}$ data analysis}
\label{sec:NH3}

The analysis of the NH$_{3}$ lines was performed using the excitation model described in \citet{Biver2012}. Since the NH$_{3}$ photodissociation lifetime is short ($\sim2~700$~s at $r_{\text{h}} = 0.7$~au)\footnote{Computed from $1/k_{\text{pd}} \times r_{\text{h}}^{2}$, where $k_{\text{pd}}$ is the photodissociation rate of NH$_{3}$ in a Solar radiation field at $1$~au and $r_{\text{h}}$ is the comet's heliocentric distance in au. Here, $k_{\text{pd}} = 1.8\times10^{-4}$~s$^{-1}$ is assumed, which is the average rate for a quiet and an active Sun from \citet{Huebner1992} and \citet{HuebnerMukherjee2015} for the NH$_{2}$(X$^{2}$B$_{1}$)~$+$~H channel, which is the dominant photodissociation channel in the Solar radiation field \citep{Heays2017}. This rate is marginally higher than the $1.5\times10^{-4}$~s$^{-1}$ value obtained in \citet{Heays2017} for the Solar radiation field at $1$~au.} and the water production rate of comet F3 is high, the excitation of NH$_{3}$ is dominated by collisional processes and its rotational levels are in thermal equilibrium. A gas kinetic temperature of $90$~K is assumed, based on constraints from close-in-date CH$_{3}$OH observations of this comet at the IRAM~$30$-m and NOEMA telescopes \citep{Biver2022b}. An expansion velocity of $1$~km~s$^{-1}$ is utilized, as derived from the line profiles observed at IRAM~$30$-m and NOEMA, which is appropriate as the field of view ($9-34\arcsec$) of these observations matches the size of the expected NH$_{3}$ coma. 

The NH$_{3}$ lines at $23-25$~GHz present a hyperfine structure with several components that are well-separated (by more than $7$~km~s$^{-1}$) from the central frequencies of the lines (as can be seen in the Cologne Database of Molecular Spectroscopy, CDMS, \citealt{Mueller2001, Mueller2005, Endres2016}). Each of these components, in turn, harbors closely spaced (by less than $0.03$~km~s$^{-1}$) quadrupole satellite lines. Two to three such satellite components contribute to the signal in the velocity interval from $-1.5$ to $1.5$~km~s$^{-1}$ that is chosen for computing the NH$_{3}$ line areas in Table~\ref{table:linearea}. The fraction of the intensity in the central hyperfine components is $50/80/89/93/96~\%$ for the $J_K = 1_1-1_1/2_2-2_2/3_3-3_3/4_4-4_4/5_5-5_5$ lines (which can be computed based on the CDMS spectroscopic entry). With this taken into consideration, the modeled line strengths show that the $J_K = 2_2-2_2$ and $3_3-3_3$ lines are, by far, the major contributors to the expected signal in this velocity interval. Based on these two lines and an RMS$^{-2}$ weighting, the resulting $3\sigma$ upper limit on the NH$_{3}$ production rate is $Q_{\text{NH}_{3}} < 1.3 \times 10^{27}$~molec~s$^{-1}$. Using the water production rate of $4.4 \times 10^{29}$~molec~s$^{-1}$ inferred from OH observations on 24 July 2020 (Section~\ref{sec:combined-OH-analysis}, an average of the \citet{Despois1981} and \citet{SchleicherAHearn1988} maser inversion models), the resulting $3\sigma$ upper limit for $Q_{\text{NH}_{3}}/Q_{\text{H}_{2}\text{O}}$ is $<0.29\%$.

\citet{Faggi2021} detected near-IR lines of NH$_{3}$ in comet F3 and derived NH$_{3}$ abundances relative to water of $0.73\pm0.10\%$ (on 31 July 2020 at $r_{\text{h}} = 0.81$~au, $Q_{\text{H}_{2}\text{O}} = (1.6\pm0.05) \times 10^{29}$~molec~s$^{-1}$) and $0.92\pm0.19\%$ (on 6 August 2020 at $r_{\text{h}} = 0.94$~au, $Q_{\text{H}_{2}\text{O}} = (1.5\pm0.2) \times 10^{29}$~molec~s$^{-1}$). The discrepancy with our value is puzzling. The measurements of NH$_{3}$ in \citet{Faggi2021} were usually performed on one or two very faint spectral lines, which prevented a direct derivation of $T_{\text{rot}}$. Consequently, it was assumed that $T_{\text{rot}}(\text{NH}_{3})=T_{\text{rot}}(\text{H}_{2}\text{O})$. Based on near-IR H$_{2}$O observations, the rotational temperature was derived to be $130$~K on 20 July and $90$~K on 31 July. If for the analysis of the GBT NH$_{3}$ lines, a $T_{\text{rot}}$ of $130$~K is assumed, then the $3\sigma$ upper limit on the NH$_{3}$ production rate is $Q_{\text{NH}_{3}} < 1.4 \times 10^{27}$~molec~s$^{-1}$ (based on the $J_K=2_2-2_2$ and $3_3-3_3$ lines and a RMS$^{-2}$ weighting). This is $\sim11\%$ higher than with a $T_{\text{rot}}$ of $90$~K; however, still not enough to explain the discrepancy with the near-IR values. Possibly, NH$_{3}$ displayed abundance variations with time. An increase in abundance relative to H$_{2}$O with increasing $r_{\text{h}}$ was observed for CH$_{3}$OH, C$_{2}$H$_{6}$, and CH$_{4}$ species in these near-IR observations, while abundances of HCN, C$_{2}$H$_{2}$, and H$_{2}$CO were stable relative to H$_{2}$O \citep{Faggi2021}. On the other hand, increases relative to H$_{2}$O with increasing $r_{\text{h}}$ were not observed for CH$_{3}$OH nor H$_{2}$CO in the IRAM~$30$-m data \citep{Biver2022b}.

It is not likely that the water production rate has been strongly overestimated in this work, thereby resulting in a low $Q_{\text{NH}_{3}}/Q_{\text{H}_{2}\text{O}}$ ratio. For the 24 July 2020 date in question ($r_{\text{h}} = 0.7$~au), here a $Q_{\text{H}_{2}\text{O}} = \left(4.42\pm0.35\right) \times 10^{29}$~molec~s$^{-1}$ is used (Section~\ref{sec:combined-OH-analysis}). Unfortunately, near-IR measurements are not available for this exact date. For 20 July 2020 ($r_{\text{h}} = 0.56$~au), \citet{Faggi2021} obtain $(4.2\pm0.2) \times 10^{29}$~molec~s$^{-1}$, whereas the $Q_{\text{H}_{2}\text{O}}$ value obtained from radio OH observation on this date is $\left(6.3\pm0.7\right) \times 10^{29}$~molec~s$^{-1}$ (weighted average of values deduced using the two maser inversion models, Table~\ref{table:lineareaNRT}). Hence, the OH-derived value is only slightly (factor of $1.5$) larger than the IR-derived value. If $Q_{\text{H}_{2}\text{O}}$ was a factor of $1.5$ lower than what has been obtained in the current analysis on 24 July 2020, the upper limit for $Q_{\text{NH}_{3}}/Q_{\text{H}_{2}\text{O}}$ would increase to $<0.44\%$, which is still well below the ratio obtained from near-IR.

The OH that is observed in comet F3 at $r_{h}=0.7$~au stems predominantly from H$_{2}$O that has exited the nucleus between $\sim4\times10^{4}$~s ($\sim12$~h, lifetime of water) and $\sim8\times10^{4}$~s ($\sim22$~h, time for the OH radicals to reach a distance corresponding to the projected GBT beam radius) earlier. On the other hand, NH$_{3}$ would be freshly released and would not survive for more than $\sim2.7\times10^{3}$~s ($\sim45$~min). There is no evidence for strong daily variability in $Q_{\text{H}_{2}\text{O}}$ in the radio observation presented nor when they are compared to the near-IR observation of (\citealt{Faggi2021}, Fig.~\ref{fig:QH2O}). Consequently, the discrepancy between near-IR and radio determinations of $Q_{\text{NH}_{3}}/Q_{\text{H}_{2}\text{O}}$ may be related to the temporal variability of $Q_{\text{NH}_{3}}$.

\begin{figure}
\includegraphics[width=\hsize]{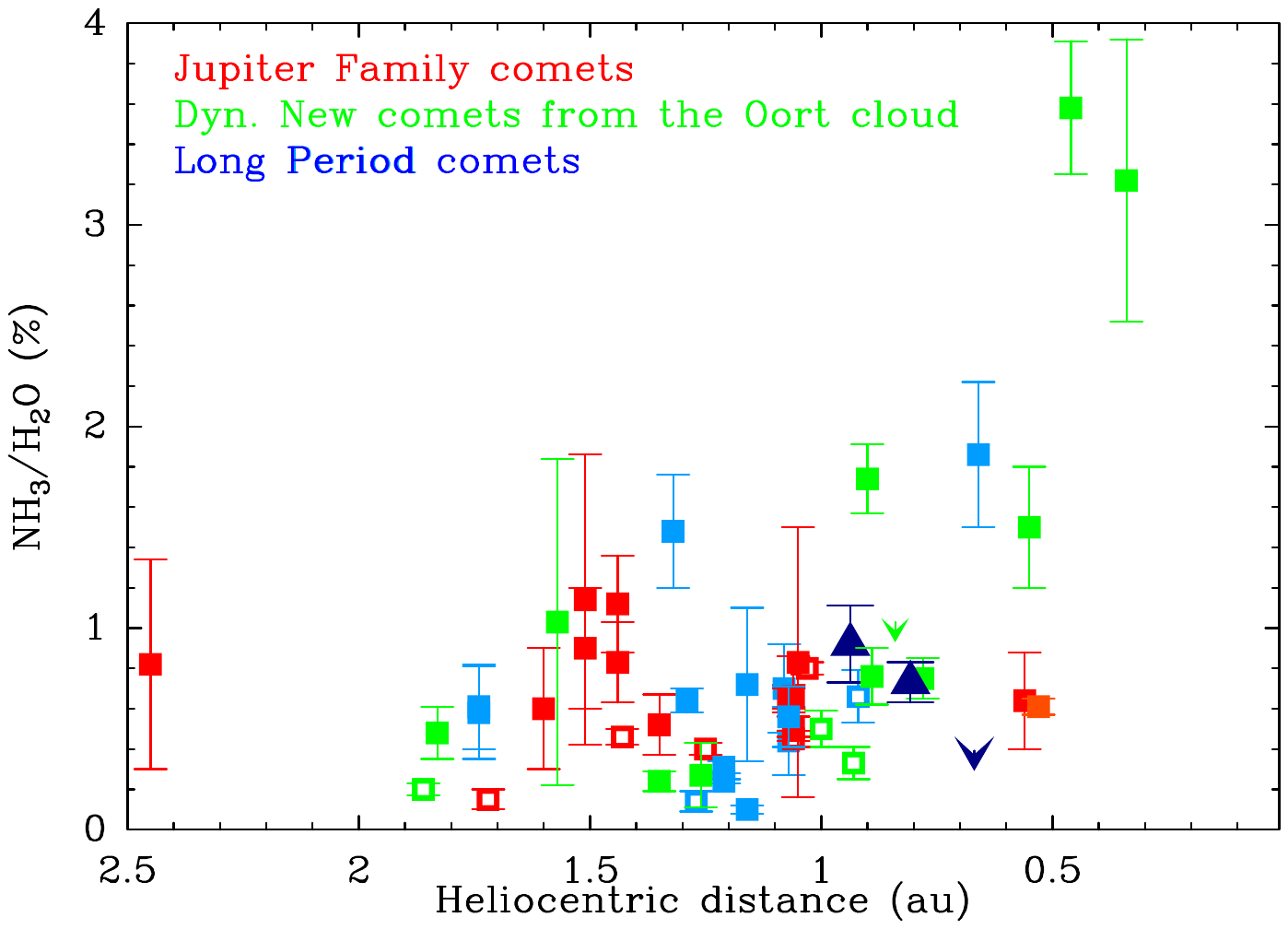} 
\caption{NH$_{3}$ abundances relative to water measured in comets from centimeter and submillimeter observations (hollow symbols), and near-IR observations (filled symbols). Different colors are used for Jupiter-family comets, dynamically new comets from the Oort Cloud, and long-period comets originating from the Oort Cloud. Downward-pointing arrowheads are $3\sigma$ upper limits. Dark blue upward-pointing triangles refer to the measurements of comet F3 from \citet{Faggi2021}. The dark blue downward-pointing arrowhead is the value derived in this work. All data used for this figure are tabulated in Tables~\ref{tabnh3comets} and~\ref{nh3_insitu} with the corresponding references.}
\label{fig:NH3-comets}
\end{figure} 


\section{Discussion}
\label{discussion}

Comet F3 was supposed to be a prime target for the investigation of the evolution of the NH$_{3}$/H$_{2}$O ratio as a function of heliocentric distance. It boasted a high water production rate ($>10^{30}$~molec~s$^{-1}$ around perihelion), which was firmly quantified during the two month-long monitoring of OH with the NRT. Its heliocentric distance at perihelion ($0.295$~au) and geocentric distance at perigee ($0.692$~au) were short. The GBT observational campaign targeted OH and NH$_{3}$ near-simultaneously, thereby ensuring robust constraints on the NH$_{3}$/H$_{2}$O ratio. However, the NH$_{3}$ hyperfine inversion lines at $23-25$~GHz eluded detection like for many other comets in the past (Section~\ref{Introduction}). The NH$_{3}$ abundance relative to water was quantified in several preceding comets (Fig.~\ref{fig:NH3-comets}) based on detections of NH$_{3}$ at $572$~GHz using \textit{Odin}, \textit{Herschel}, and \textit{Rosetta}/MIRO (\citealt{Biver2007b, Biver2012, Biver2019}, and unpublished results from N. Biver, personal communication) and in the near-IR (\citealt{DelloRusso2016b, Lippi2021}, and other references in Table~\ref{tabnh3comets}). The upper limit obtained for comet F3 (blue downward-pointing arrowhead in Fig.~\ref{fig:NH3-comets}) is in the low range of values measured in comets at $0.7$~au from the Sun.

There is a trend for higher NH$_{3}$ abundances at low ($<1$~au) heliocentric distances (Fig.~\ref{fig:NH3-comets}), suggesting a possible contribution of NH$_{3}$ released from the thermal degradation of compounds on grains at small $r_{\rm h}$. This is supported by the radial profiles of NH$_{3}$, which are often more extended than expected for a nuclear source of NH$_{3}$ \citep[e.g.,][]{DiSanti2016, DelloRusso2022}. A similar increase at low heliocentric distances is also seen in the $Q(\text{NH}_{3})/Q(\text{C}_{2}\text{H}_{6})$ ratio (fig.~$5$ of \citealt{Mumma2019}). The thermal decomposition of ammonium salts discovered in comet 67P/C-G \citep{Quirico2016, Altwegg2020a, Poch2020} has been proposed to explain the excess of NH$_{3}$ production at small $r_{\rm h}$ \citep{Mumma2019}. This excess of production is not observed for comet F3, based on the here-obtained NH$_{3}$/H$_{2}$O upper limit and the NH$_{3}$ abundance determined in the near-IR \citep{Faggi2021}. A contributing factor could be the amount of water-poor dust launched into the coma of a comet. At $0.7$~au from the Sun, the equilibrium temperature of dust grains is already above the $\sim160-230$~K threshold for thermal degradation of some ammonium salts. The increase in the NH$_{3}$/H$_{2}$O ratio must therefore be correlated with a higher dust density in the coma. Furthermore, given enough dust in the coma, an initial grain size distribution can be altered by fragmentation as it moves outward through the coma, which may lead to an increase in the small grain population as a function of distance from the nucleus. The small grains can be super-heated above the equilibrium temperature; whereby, salts with the highest binding energies would also thermally disintegrate.
This dust must be dry, i.e., water-poor, because otherwise the amount of H$_{2}$O would increase together with NH$_{3}$ and potentially mask the increase in the relative abundance of NH$_{3}$.

The dust-to-gas ratio in a cometary coma may vary drastically with outbursts. Comet F3 was not observed to undergo outbursts during its 2022 perihelion passage, but rather to only display strong jet activity (\citealt{Combi2021}, Section~\ref{Introduction}). Additional observations of NH$_{3}$ at low $r_{\rm h}$ are obviously needed. Although, it is not excluded that enhanced NH$_{3}$/H$_{2}$O ratios may be short-lived as ammonium salts may not survive for long on dust in the coma following an injection of dust from an outburst. Typically, outbursts are more frequent at smaller heliocentric distances, in agreement with the NH$_{3}$/H$_{2}$O increasing trend. Continuous monitoring pre- and post-outburst would allow the lifetime of ammonium salts in the coma to be estimated.

Several salts have been identified to be present on the surface of 67P/C-G: NH$^{+}_{4}$Cl$^{-}$, NH$^{+}_{4}$CN$^{-}$, NH$^{+}_{4}$OCN$^{-}$, NH$^{+}_{4}$HCOO$^{-}$, NH$^{+}_{4}$CH$_{3}$COO$^{-}$, NH$^{+}_{4}$SH$^{-}$, and NH$^{+}_{4}$F$^{-}$ \citep{Altwegg2020a, Altwegg2022}. As ammonium salts degrade at higher temperatures at smaller heliocentric distances, not only the NH$_{3}$/H$_{2}$O ratio should display an increasing trend, but also the ratio of the acid-counterparts (HCl, HCN, HOCN, HCOOH, HCOOCH$_{3}$) relative to H$_{2}$O. However, to what extent each of these individual salts makes a significant contribution relative to the amount of these species already present in the ice in the nucleus is not clear. Modeling work is required to quantify these effects. For the case of the CN radical, it has been shown that it does appear to have a distributed source (e.g., \citealt{Opitom2016}) and requires an additional parent molecule to match its signal strength in the ROSINA measurements of the inner coma of 67P/C-G \citep{Haenni2020, Haenni2021}, thus likely being a product of ammonium salt thermal degradation. This has also been observed for NH$_{2}$, a photodissociation product of NH$_{3}$ \citep{Opitom2019b}. An important step forward would be the comparison of the $^{14}$N/$^{15}$N isotopic ratio in the dust with that in various N-bearing volatiles. It has been shown that the $^{14}$N/$^{15}$N isotopic ratio for NH$_{3}$, NO, N$_{2}$ in 67P/C-G is in the $93-160$ range, which is consistent with the $\sim140$ measured in HCN, CN, and NH$_{2}$ in other comets \citep{Altwegg2019, Bockelee-Morvan2015a, Biver2022a}. This ratio has not been reported for the dust of 67P/C-G.


\section{Conclusions}
\label{conclusions}

This paper presents the two month-long monitoring campaign of the OH emission from comet C/2020 F3 (NEOWISE) at the Nan\c{c}ay Radio Telescope, which is used to determine the H$_{2}$O production rate. Furthermore, GBT observations of F3 targeting OH lines on two separate days (24 July and 11 August 2020) and the NH$_{3}$ line observations contemporaneous with the first date are presented. The main results are as follows.
\begin{enumerate}
	\item The OH parent expansion velocity ($v_{p}$) increases from $1.0$ to $1.8$~km~s$^{-1}$ with decreasing heliocentric distance from $0.72$ to $0.52$~au.
	\item The OH quenching radius ($r_{q}$) is determined to be $\left(5.96\pm0.10\right)\times10^{4}$~km on 24 July 2020 ($r_{\text{h}} = 0.7$~au) from the analysis of the concurrent NRT and GBT OH observations. This value consistently explains the OH line intensities measured by the GBT and NRT for an OH production rate ($Q_{\text{OH}}$) of $\left(4.02\pm0.32\right)\times10^{29}$~molec~s$^{-1}$. This $r_{q}$ value is consistent with the \citet{Gerard1990} prescription.
	\item The Greenstein effect \citep{Greenstein1958} is observationally demonstrated in the OH observations of 11 August 2020 taken at the GBT, yielding a narrow, strongly blueshifted line in absorption. The signal-to-noise ratio of the data is not high enough to distinguish between the \citet{Despois1981} and \citet{SchleicherAHearn1988} maser inversion models.
	\item One day before perihelion (2 July 2020), the H$_{2}$O production rate ($Q_{\text{H}_{2}\text{O}}$) was very high ($\sim2\times10^{30}$~molec~s$^{-1}$) in comet F3. Pre-perihelion, $Q_{\text{H}_{2}\text{O}}$ increases with decreasing heliocentric distance and agrees within a factor of $2$ with Ly-$\alpha$ observations using SOHO/SWAN of \citet{Combi2021}. Post-perihelion, $Q_{\text{H}_{2}\text{O}}$ decreases with the increasing heliocentric distance and is in excellent agreement with Ly-$\alpha$ observations.
	\item The $3\sigma$ upper limit for $Q_{\text{NH}_{3}}/Q_{\text{H}_{2}\text{O}}$ is $<0.29\%$ for comet F3 at $0.7$~au from the Sun (post-perihelion on 24 July 2020), which is in the low range of values obtained for other comets at similar heliocentric distances.
	\item The differences in the $\text{NH}_{3}/\text{H}_{2}\text{O}$ ratios measured for comet F3 with radio and near-IR observations may hint at this ratio being highly variable with time in a cometary coma.
\end{enumerate}

\vspace{0.5cm}

\begin{acknowledgements}
This work is supported by the Swiss National Science Foundation (SNSF) Ambizione grant no. 180079, the Center for Space and Habitability (CSH) Fellowship, and the IAU Gruber Foundation Fellowship. MAC, SBC, and SNM's work was supported by the NASA Planetary Science Division Internal Scientist Funding Program through the Fundamental Laboratory Research (FLaRe) work package. This work benefited from discussions held with the international team \#461 ``Provenances of our Solar System's Relics'' (team leaders Maria N. Drozdovskaya and Cyrielle Opitom) at the International Space Science Institute, Bern, Switzerland.

The Green Bank Observatory is a facility of the National Science Foundation operated under cooperative agreement by Associated Universities, Inc. The authors are grateful for the guidance from Tapasi Ghosh and project friend Larry Morgan in reducing L-band and KFPA data, respectively. The Nan\c{c}ay Radio Observatory is operated by the Paris Observatory, associated with the French Centre National de la Recherche Scientifique (CNRS) and with the University of Orl\'eans. The authors thank the referee, Mike Mumma, for constructive comments that have strengthened the manuscript.
\end{acknowledgements}

\bibliographystyle{aa} 
\bibliography{Neowise} 


\begin{appendix}
\section{Averaged NRT spectra of comet F3 for 18-27 July 2020}
\label{apx:aveNRT}

\begin{figure}[hbt!]
\includegraphics[width=\hsize]{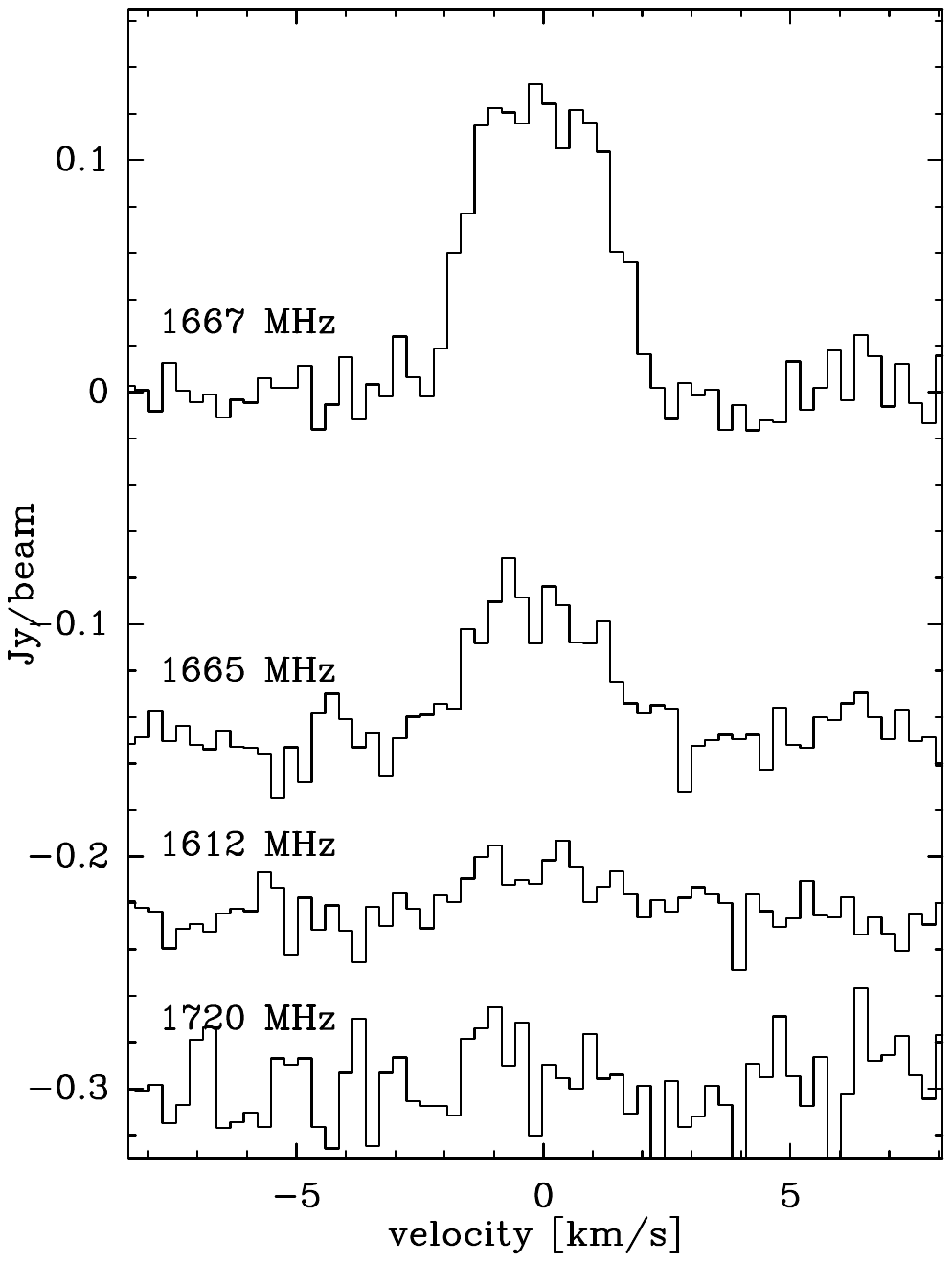} 
\caption{Four OH $18$-cm lines observed in comet F3 with the NRT averaged from 18 to 27 July 2020. The expected statistical relative intensities of the $1~667:1~665:1~612:1~720$~MHz lines are $9:5:1:1$, which is in good agreement with the displayed observations.}
\label{fig:aveNRT}
\end{figure} 

\section{Ammonia in comets}
\label{apx:tableNH3}
NH$_{3}$ abundances relative to water measured in comets from ground-based radio and IR observations are tabulated in Table~\ref{tabnh3comets}, while those obtained with in situ measurements are given in Table~\ref{nh3_insitu} with the corresponding references. For Fig.~\ref{fig:NH3-comets}, all ground-based values have been used from Table~\ref{tabnh3comets} except for the measurement for comet C/1995~O1 (Hale-Bopp) from \citet{Hirota1999} due to its large, unconstraining error bars. From Table~\ref{nh3_insitu}, only the two reported in lines two and three have been used for Fig.~\ref{fig:NH3-comets}. The other measurements for comet 67P/Churyumov-Gerasimenko have been tabulated in order to illustrate that a range of values has been measured at this comet over the 2-year duration of the ESA \textit{Rosetta} mission with two independent instruments: MIRO and ROSINA. The data point of 1P/Halley has been excluded from Fig.~\ref{fig:NH3-comets}, because its error bars are large and do not allow for a meaningful comparison.

\begin{table*}[hbt!]
\caption[]{Abundances of ammonia in comets from ground-based observations.}
\label{tabnh3comets}
\begin{center}
\begin{tabular}{llllll}
\hline\hline
UT date & $r_{h}$ & $Q_{\rm NH_3}/Q_{\rm H_2O}$ & \multicolumn{2}{l}{Comet} & Reference\\
$($yyyy/mm/dd.d) & (au) & ($\%$) & Type\tablefootmark{a}  & Name &\\
\hline
\multicolumn{6}{l}{Radio data ($\lambda\sim$cm)\tablefootmark{b}}\\
\hline
1996 03 24   & 1.07 & $0.44\pm0.17$ & LPC & C/1996~B2 (Hyakutake) & \citet{Wootten1996, Wootten1996c6362, Palmer1996}\\
             &      &               &     &                       & (A. Wootten \& B. Butler, pers. comm.)\tablefootmark{c}\\
1997 03 30   & 0.92 & $0.66\pm0.13$ & LPC & C/1995~O1 (Hale-Bopp) & \citet{Bird1997}\tablefootmark{d}\\
1997 04 20   & 0.98 & $1.8\pm0.9$   & LPC & C/1995~O1 (Hale-Bopp) & \citet{Hirota1999}\\
1997 05 22   & 1.27 & $0.14\pm0.05$ & LPC & C/1995~O1 (Hale-Bopp) & \citet{Butler2002}\\
             &      &               &     &                       & (A. Wootten \& B. Butler, pers. comm.)\tablefootmark{e}\\
2020 07 24.8 & 0.67 & $<0.29$       & LPC & C/2020~F3 (NEOWISE)   & This paper\\
\hline
\multicolumn{6}{l}{Radio data ($\lambda=0.52$~mm)}\\
\hline
2004 04 29.8 & 1.00 & $0.50\pm0.09$ & DNC & C/2001~Q4 (NEAT)      & \citet{Biver2007b}\\
2004 05 25.9 & 0.93 & $0.33\pm0.08$ & DNC & C/2002~T7 (LINEAR)    & \citet{Biver2007b}\\
2010 07 19.1 & 1.43 & $0.46\pm0.04$ & JFC & 10P/Tempel 2          & \citet{Biver2012}\\
2010 10 30.6 & 1.06 & $0.45\pm0.04$ & JFC & 103P/Hartley 2        & Unpublished results from \textit{Herschel}\\
             &      &               &     &                       & (N. Biver, pers. comm.)\\
2011 08 14.0 & 1.03 & $0.80\pm0.03$ & JFC & 45P/Honda-Mrkos-Pajdu\v{s}\'{a}kov\'{a} & idem\\
2011 10 08.9 & 1.86 & $0.20\pm0.03$ & DNC & C/2009~P1 (Garradd) & idem\\
\hline
\multicolumn{6}{l}{Infrared data}\\
\hline
2007 12 23   & 1.16 & $0.72\pm0.38$ & HTC & 8P/Tuttle & \citet{Lippi2021}\\
2007 10 29   & 2.45 & $0.82\pm0.52$ & JFC & 17P/Holmes & idem\\
2005 07 04 & 1.51 & $1.14\pm0.72$ & JFC & 9P/Tempel 1 & idem\\
2010 07 26 & 1.44 & $1.12\pm0.24$ & JFC & 10P/Tempel 2 & idem\\
2010 10 30 & 1.07 & $0.64\pm0.06$ & JFC & 103P/Hartley 2 & idem\\
1999 08 19 & 1.07 & $0.56\pm0.15$ & LPC & C/1999~H1 (Lee) & idem\\
2005 01 19 & 1.21 & $0.24\pm0.01$ & LPC & C/2004~Q2 (Machholz) & idem\\
2009 02 01 & 1.26 & $0.27\pm0.16$ & DNC & C/2007~N3 (Lulin) & idem\\
2008 07 09 & 0.89 & $0.76\pm0.14$ & DNC & C/2007~W1 (Boattini) & idem\\
2012 01 09 & 1.57 & $1.03\pm0.81$ & DNC & C/2009~P1 (Garradd) & idem\\
2013 06 20 & 1.74 & $0.58\pm0.23$ & LPC & C/2012~F6 (Lemmon) & idem\\
2013 10 25 & 1.32 & $1.48\pm0.28$ & LPC & C/2013~R1 (Lovejoy) & idem\\
2008 08 11   & 1.35 & $0.52\pm0.15$ & JFC &   6P/d'Arrest        & \citet{DelloRusso2016b}\\
2005 07 04   & 1.51 & $0.9 \pm0.3 $ & JFC &   9P/Tempel 1    & idem\\
2010 07 26   & 1.44 & $0.83\pm0.2 $ & JFC &  10P/Tempel 2    & idem\\
2010 02 23   & 1.60 & $0.6 \pm0.3 $ & JFC &  81P/Wild 2    & idem\\
2010 10 30   & 1.07 & $0.66\pm0.06$ & JFC & 103P/Hartley 2    & idem\\
1999 08 21   & 1.08 & $0.70\pm0.22$ & LPC & C/1999~H1 (Lee)    & idem\\
2005 01 25   & 1.21 & $0.31\pm0.03$ & LPC & C/2004~Q2 (Machholz)    & idem\\
2007 01 28   & 0.55 & $1.5 \pm0.3 $ & DNC & C/2006~P1 (McNaught)    & idem\\
2009 02 15   & 1.35 & $0.24\pm0.05$ & DNC & C/2007~N3 (Lulin)    & idem\\
2008 07 10   & 0.90 & $1.74\pm0.17$ & DNC & C/2007~W1 (Boattini)    & idem\\
2011 10 13   & 1.83 & $0.48\pm0.13$ & DNC & C/2009~P1 (Garradd)    & idem\\
2013 06 20   & 1.74 & $0.61\pm0.21$ & LPC & C/2012~F6 (Lemmon)    & idem\\
2013 11 07.6 & 0.84 & $<0.93$       & DNC & C/2012~S1 (ISON) & \citet{DiSanti2016}\\
2013 11 19.8 & 0.46 & $3.58\pm0.33$ & DNC & C/2012~S1 (ISON) & idem\\
2013 11 23.0 & 0.34 & $3.22\pm0.70$ & DNC & C/2012~S1 (ISON) & idem\\
2013 11 07   & 1.16 & $0.10\pm0.02$ & LPC & C/2013~R1 (Lovejoy)    & \citet{DelloRusso2016b}\\
2014 02 03   & 1.29 & $0.64\pm0.06$ & LPC & C/2014~Q2 (Lovejoy) & \citet{DelloRusso2022}\\
2017 01 08   & 0.56 & $0.64\pm0.24$ & JFC & 45P/Honda-Mrkos-Pajdu\v{s}\'{a}kov\'{a} & \citet{DiSanti2017}\\
2017 02 16   & 1.05 & $0.83\pm0.67$ & JFC & 45P/Honda-Mrkos-Pajdu\v{s}\'{a}kov\'{a} & \citet{DelloRusso2020}\\
2017 03 25   & 0.53 & $0.61\pm0.04$ & ETC & 2P/Encke & \citet{Roth2018}\\
2014 09 05   & 0.78 & $0.75\pm0.10$ & DNC & C/2013~V5 (Oukaimeden) & \citet{DiSanti2018}\\
2017 04 04   & 0.66 & $1.86\pm0.36$ & LPC & C/2017~E4 (Lovejoy) & \citet{Faggi2018}\\
2018 12 17   & 1.06 & $0.66\pm0.20$ & JFC & 46P/Wirtanen & \citet{Bonev2021}\\
2018 12 21   & 1.06 & $0.50\pm0.06$ & JFC & 46P/Wirtanen & \citet{Khan2021}\\
2020 07 31.1 & 0.81 & $0.73\pm0.10$ & LPC & C/2020~F3 (NEOWISE) & \citet{Faggi2021}\\
2020 08 06.3 & 0.94 & $0.92\pm0.19$ & LPC & C/2020~F3 (NEOWISE) & idem\\
\hline
\end{tabular}
\end{center}
\tablefoot{
  \tablefoottext{a}{ETC = Encke-Type Comet, JFC = Jupiter-Family Comet,
    LPC = Long-Period Comets originating from the Oort Cloud, DNC = Dynamically New Oort Cloud Comet, HTC = Halley-Type Comet.}\\
  \tablefoottext{b}{For other tentative detections and non-detections in further comets, see also \citet{Altenhoff1983, Bird1987, Churchwell1976, Bird2002, Hatchell2005}.}\\
	\tablefoottext{c}{$Q_{\text{NH}_{3}}=\left(7.4\pm2.6\right)\times10^{26}$~molec~s$^{-1}$ from the tabulated references and $Q_{\text{H}_{2}\text{O}}=\left(1.7\pm0.3\right)\times10^{29}$~molec~s$^{-1}$ from \citet{Mumma1996a}.}\\
  \tablefoottext{d}{$Q_{\text{NH}_{3}}=6.6\times10^{28}$~molec~s$^{-1}$ from \citet{Bird1997} and $Q_{\text{H}_{2}\text{O}}=1\times10^{31}$~molec~s$^{-1}$, which is the best-possible estimate for the dates of NH$_{3}$ observations based on SOHO/SWAN measurements of \citet{Combi2002}, NRT data analysis of \citet{Colom1999}, and near-ultraviolet OH observations of \citet{Harris2002b}.}\\
	\tablefoottext{e}{$Q_{\text{NH}_{3}}=\left(7.9\pm2.8\right)\times10^{27}$~molec~s$^{-1}$ from the tabulated references and $Q_{\text{H}_{2}\text{O}}=\left(5.6\pm0.3\right)\times10^{30}$~molec~s$^{-1}$ \citet{Combi2000}.}}
\end{table*}

\begin{sidewaystable*}
\caption{Abundances of ammonia in comets from in situ measurements.}
\label{nh3_insitu}
\begin{center}
\begin{tabular}{lllllll}
\hline
\hline
UT date & $r_{h}$ & Instrument & $Q_{\rm NH_3}/Q_{\rm H_2O}$ & \multicolumn{2}{l}{Comet\tablefootmark{a}} & Reference \\
$($yyyy/mm/dd.d) & (au) & & ($\%$)\tablefootmark{b} & Type & Name &  \\
\hline
1986 03 14   & 0.90                   & \textit{Giotto}/NMS\tablefootmark{c}   & $1.5\pm0.6$   & HTC & 1P/Halley & \citet{Meier1994, Rubin2011}\\
2015 08 15.0 & 1.25 (pre-perihelion)  & \textit{Rosetta}/MIRO                  & $0.40\pm0.03$ & JFC & 67P/C-G   &  \citet{Biver2019}\\  
2015 11 22.2 & 1.72 (post-perihelion) & \textit{Rosetta}/MIRO                  & $0.15\pm0.05$ & JFC & 67P/C-G   &  idem\\
\hline
& $3.2$             & \textit{Rosetta}/ROSINA & $0.06^{\text{N summer}}-0.15^{\text{S winter}}$     & JFC & 67P/C-G & \citet{LeRoy2015}\\
& (pre-perihelion, pre-in-equinox)   & \textit{Rosetta}/MIRO & $0.2$                                & JFC & 67P/C-G & \citet{Biver2019}\\
\hline
& $2.0-2.7$         & \textit{Rosetta}/ROSINA & $0.0378\pm0.0013$                                   & JFC & 67P/C-G & \citet{Gasc2017}\\
& (post-perihelion, pre-out-equinox) & \textit{Rosetta}/MIRO & $0.1$                                & JFC & 67P/C-G & \citet{Biver2019}\\
\hline
& $3.1-3.5$         & \textit{Rosetta}/ROSINA & $0.0132\pm0.0013$                                   & JFC & 67P/C-G & \citet{Gasc2017}\\
& (post-perihelion, post-out-equinox) & & & & & \\
\hline
& Bulk (May 2015)   & \textit{Rosetta}/ROSINA & $0.67\pm0.20^{\text{S summer}}$                     & JFC & 67P/C-G & \citet{Rubin2019a}\\
& (pre-perihelion, post-in-equinox) & & & & & \\
\hline
& 2-year integrated & \textit{Rosetta}/ROSINA & $(0.34\pm0.11)^{\text{S}}-(0.43\pm0.13)^{\text{N}}$ & JFC & 67P/C-G & \citet{Laeuter2022}\\
&                   & \textit{Rosetta}/MIRO   & $0.34\pm0.01$                                       & JFC & 67P/C-G\\
\hline
\end{tabular}
\end{center}
\tablefoot{
  \tablefoottext{a}{HTC = Halley-Type Comet, JFC = Jupiter-Family Comet, 67P/C-G = 67P/Churyumov-Gerasimenko.} \tablefoottext{b}{N and S denote the northern and southern hemispheres of 67P/Churyumov-Gerasimenko, respectively.} \tablefoottext{c}{NMS = Neutral Mass Spectrometer.}}
\end{sidewaystable*}

\end{appendix}


\end{document}